\documentclass[aps,prb,twocolumn,epsf,epsfig,amsmath]{revtex4}
\usepackage{latexsym}
\usepackage{hyperref}
\usepackage{times}
\usepackage{graphicx}
\usepackage{amsmath}
\usepackage{dcolumn}
\usepackage{bm}
\usepackage{textcomp}
\usepackage{color}
\usepackage{amssymb}
\usepackage{xcolor}
\usepackage{float}
\usepackage{array}
\usepackage[export]{adjustbox}

\begin{document}
\title{Self induced Hall Effect in current carrying bar}
\author{M. V. Cheremisin}
\affiliation{A.F.Ioffe Physical-Technical Institute, St.Petersburg, Russia}
\date{\today}
\begin{abstract}
The longitudinal current in a three-dimensional conductor is accompanied by transverse magnetic field in a specimen bulk.
The absence of the transverse current in a sample bulk requires a nonzero Hall electric field in transverse cross-section which provides the Lorentz
force cancelation. The longitudinal current itself can be viewed as a collective drift of carriers in crossed magnetic and electric(Hall) fields.
At low temperatures the enhanced carrier viscosity leads to nonuniform current flow whose transverse profile is sensitive to presence of
$collinear$ diamagnetic currents at the sample inner walls. Former dissipative the longitudinal current becomes purely diamagnetic
at certain critical temperature. The superconductivity sets on. The current and transverse magnetic field become pushed out
from the sample bulk towards the inner wall. Magnetic properties of a sample resemble those expected for ideal diamagnet. The
threshold of superconductivity is calculated for arbitrary temperature, disorder strength, sample size and current and(or) magnetic
field strength. Sample-size and magnetic field driven transition from normal metal phase to superconductivity has been studied.
The superconductivity phase is not discussed.
\end{abstract}
\maketitle

\section{Introduction}
\label{Introduction}
Usually, the Hall measurements\cite{Hall1879} imply a presence of external magnetic field affecting the carrier
transport. The typical scheme of the Hall bar geometry sample is shown in Fig.\ref{Fig1}. Evidence shows that the current itself
may produce a finite magnetic field which, in turn, influences the current carrying state.
In present paper, we take the interest in a special case of the magnetic field induced by the current itself. We
reveal a self-consistent Hall Effect. Taking into account finite carrier viscosity and diamagnetic
currents persistent at inner sample walls we demonstrate feasibility of the superconductivity
at low temperatures.

\section{Current induced Hall Effect in a bar sample}
\label{Current induced Hall Effect in 2D bar}

The conventional Drude equation for electron gas in presence of arbitrary electric $\mathbf{E}$ and magnetic $\mathbf{B}$ fields yields
\begin{equation}
\frac{\partial \mathbf{V}}{\partial t}=\frac{e \mathbf{E}}{m}+\left [\mathbf{V}\times \mathbf{\Omega}_{c} \right]-\frac{\mathbf{V}}{\tau},
\label{Drude}\\
\end{equation}
where $\mathbf{V}$ is the electron flux velocity, $e$ is the absolute value of the electronic charge. Then, $\mathbf{\Omega}_{c}=\frac{e\mathbf{B}}{mc}$
is a vector whose absolute value corresponds to local cyclotron frequency, $m$ is the effective mass, $\tau$ is the momentum relaxation
time due to collisions with impurities and(or) phonons.

\begin{figure}[tbp]
\begin{center}\leavevmode
\includegraphics[width=1.0\linewidth]{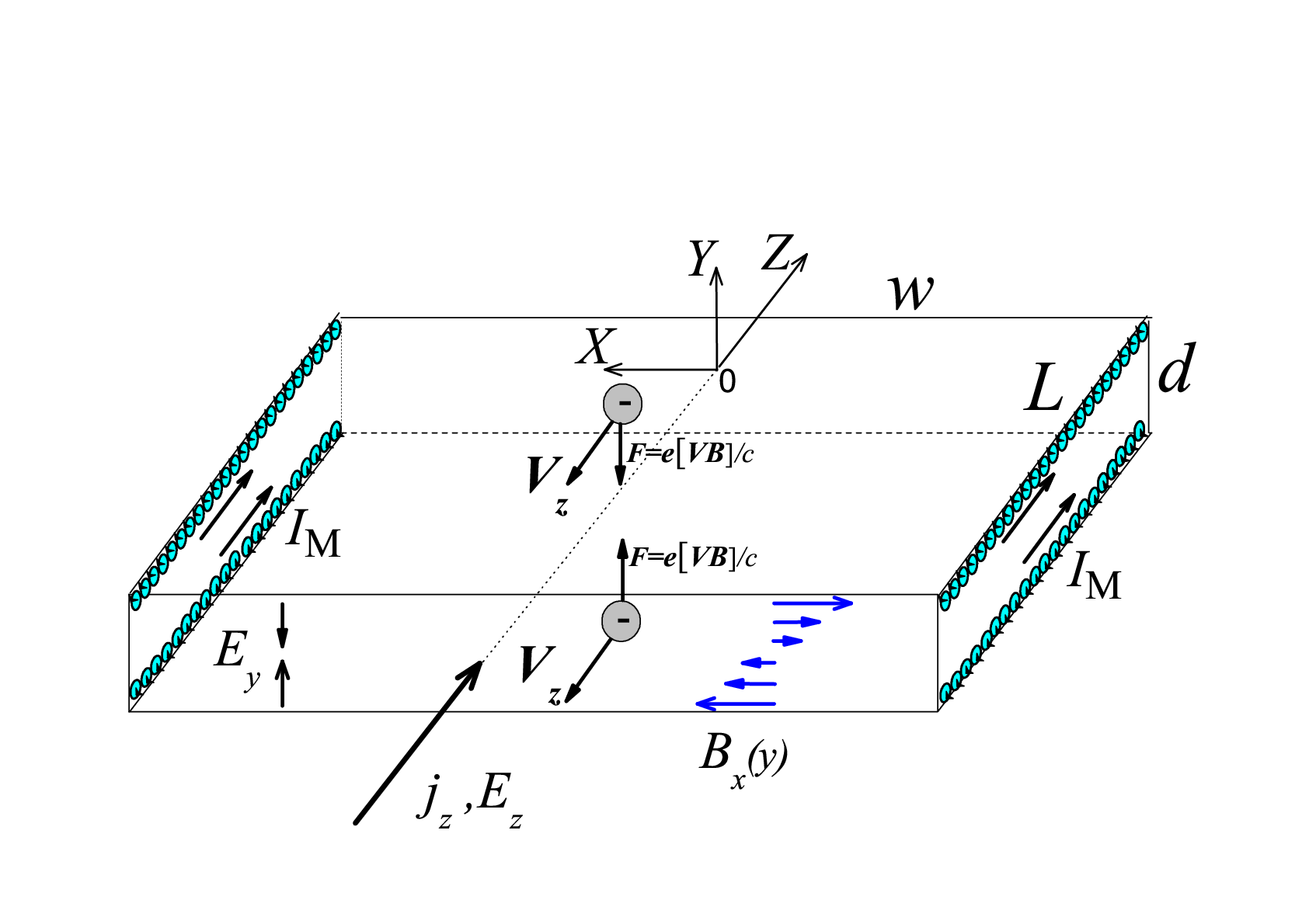} \caption[]{\label{Fig1} Schematic view of the self-consistent Hall effect. The Lorentz force is shown for electron moved in top(bottom) half-space of the sample bulk. The average diamagnetic currents $I_{M}$ caused by microscopic cyclotron movement of electrons in the vicinity of the top(bottom) inner wall are shown.}
\end{center}
\end{figure}

For steady state Eq.(\ref{Drude}) yields
\begin{equation}
\mathbf{V}=\mu \mathbf{E}+\left [\mathbf{V} \times \mathbf{\Omega}_{c}\tau \right],
\label{vector equation}\\
\end{equation}
where $\mu=\frac{e\tau}{m}$ is the carrier mobility. For arbitrary orientation of the electric
and the magnetic fields the exact solution of Eq.(\ref{vector equation}) is straightforward \cite{Anselm78}.

We further restrict ourself to a certain sample geometry, namely consider the bar sample of thickness $d$ which is less than both the width $w$
and the length $L$. Let a voltage source(not shown in Fig.\ref{Fig1}) is attached to sample contacts providing the longitudinal
electric field $E_{z}$ in a sample bulk. Evidence shows that longitudinal carrier velocity $V_{z}$ and, hence, the respective current
density $j_{z}=neV_{z}$ are uniform for constant electron density $n$. One may easily find x-component of the transverse magnetic field $B_{x}(Y)=-\frac{4\pi j_{z}}{c}Y$ followed from Biot-Savart law. Notably, the transverse magnetic field reaches its maximal value
$B_{0}=\frac{2\pi I}{cw}$ at the top(bottom) wall of the slab, where $I=j_{z}dw$ is the total current. Evidence shows that the
transverse components of the current density are absent in a sample bulk, i.e $j_{x,y}\equiv 0$. Hence, in presence of the transverse magnetic field $B_{x}$ a
nonzero y-component of the electric field $E_{y}$ must exist to prevent Lorentz force
action $\sim V_{z}B_{x}/c$ Evidently, the build-in transverse field $E_{y}$ plays a role of the Hall electric field in
conventional description\cite{Hall1879}.

Following the above reasoning we re-write Eq.(\ref{vector equation}) for both the longitudinal $V_{z}$ and transverse $V_{y}=0$ components
of the carrier velocity:
\begin{eqnarray}
V_{z}=\mu E_{z},
\label{velocity_Drude}\\
V_{z}=c\frac{E_{y}}{B_{x}}.
\label{velocity_Drude+}
\end{eqnarray}
Eq.(\ref{velocity_Drude}) represents a familiar description of differential Ohm's law $j_{z}=j_{D}$, where $j_{D}=ne\mu E_{z}$
is the Drude current. By contrast, Eq.(\ref{velocity_Drude+}) presents the novel view on the longitudinal current as a drift of
carriers in crossed $E_{y}\perp B_{x}$ fields, i.e.
ascribes self-consistent Hall Effect. We argue that the build-in electric field $E_{y}=4\pi neY\left(\frac{V_{z}}{c}\right)^{2}$
defines volumetric charge density $Q=\text{div}\mathbf{E}/4\pi=ne\left(\frac{V_{z}}{c}\right)^{2}$\cite{Matzek68,McDonald2010}. Thus,
a sample bulk is charged, i.e. $Q/ne\ll 1$.

\section{Hall Effect: Nonuniform viscous flow}
\label{Nonuniform current flow}
We now intend to answer a question whether the current carrying state in a bar could be nonuniform in
transverse direction, namely we presume $V_{z}(Y)$. To resolve the problem, we will use Navier-Stokes equation valid for viscous electron fluid
\begin{equation}
\frac{\partial \mathbf{V}}{\partial t}+(\mathbf{V}\nabla)\mathbf{V}=\frac{e \mathbf{E}}{m}+\left [\mathbf{V}\times \mathbf{\Omega}_{c} \right]+\hat{\eta} \triangle \mathbf{V}-\frac{\mathbf{V}}{\tau}
\label{Navier-Stokes}\
\end{equation}
in presence of the magnetic field. Here, $\hat{\eta}$ is the viscosity tensor\cite{Steinberg58} whose
longitudinal and transverse components

\begin{eqnarray}
\eta_{xx}=\eta_{yy}=\frac{\eta}{1+4\Omega_{c}^{2}\tau_{\eta}^{2}},
\label{viscosity_components}\\
\eta_{xy}=-\eta_{yx}=\eta_{xx}2\Omega_{c}\tau_{\eta}
\nonumber
\end{eqnarray}
depend on magnetic field. Then, $\eta=\frac{1}{5} V^{2}_{F}\tau_{\eta}$ is the kinematic viscosity of the carriers at zero magnetic field, $V_{F}$ is the Fermi velocity, $\tau_{\eta}$ denotes viscosity relaxation time whose exact form will be specified further. Viscosity effects start to be important\cite{Gurzhi63,Dyakonov93} when the mean free path $l_{\eta}=V_{F}\tau_{\eta}$ becomes less and(or) comparable to that $l=V_{F}\tau$ caused by phonons and(or) impurities and typical length scale of the sample.

One may easily check that for present case of nonuniform flow the Euler term in a left part of Eq.(\ref{Navier-Stokes}) is equal to zero. Therefore,
for steady state Eq.(\ref{Navier-Stokes}) can be re-written for both the longitudinal and transverse direction as it follows
\begin{eqnarray}
\eta_{xx} \tau \frac{ \partial^{2} V_{z}}{\partial^{2} Y}-V_{z}+\mu E_{z}=0,
\label{XX_velocity_Stokes}\\
\eta_{yx} \frac{ \partial^{2} V_{z}}{\partial^{2} Y}+\frac{eE_{y}}{m}-\frac{V_{z}eB_{x}}{mc}=0.
\label{YX_velocity_Stokes}\\
\nonumber
\end{eqnarray}
Our primary interest concerns Eq.(\ref{XX_velocity_Stokes}) which determines the nonuniform velocity profile $V_{z}(Y)$
and, in turn, the transverse magnetic field $B_{x}(Y)$
\begin{equation}
B_{x}=-\frac{2\pi ne}{c}\int\limits^{Y}_{-Y}V_{z}(Y) dY.
\label{Transversed_Field}\
\end{equation}
Introducing the dimensionless velocity $v=V_{z}/\mu E_{z}$ and the reduced transverse co-ordinate $y=Y/d$, one may rewrite
Eq.(\ref{XX_velocity_Stokes}) as it follows
\begin{equation}
\frac{\eta_{xx}}{\eta} \nu^{-2} \frac{ \partial^{2} v}{\partial^{2} y}-v+1=0.
\label{Vz_Stokes_Modified}\\
\end{equation}
where $\nu=d/\lambda$ is the dimensionless parameter. Then, $\lambda=\sqrt{\eta \tau}$ denotes the typical length scale of
velocity profile $v_{z}(y)$ sharpness. The condition $\nu \ll 1$( $\nu \gg 1$ ) determines the high(low)-viscous electron gas respectively.

We argue that solving of Eq.(\ref{Vz_Stokes_Modified}) is complicated because of field dependent viscosity pre-factor $\eta_{xx}(\Omega_{c})\rightarrow \eta_{xx}(y)$ in the highest derivative term. In principle, Eq.(\ref{Vz_Stokes_Modified}) can be expressed in terms of the reduced magnetic field $B_{x}(y)/B_{0}$ via relationship $v=\frac{1}{2B_{0}}\frac{dB_{x}}{dy}$ but still remains difficult for analytic analysis. However, we may restrict ourself to low-current and(or) small magnetic field case when $\Omega_{c}\tau_{\eta}\ll 1$. By doing so, the longitudinal viscosity can be kept constant $\eta_{xx}\sim \eta$. The solution of Eq.(\ref{XX_velocity_Stokes}) becomes straightforward:
\begin{equation}
v(y)=1+C_{1}\cosh(\nu y)+C_{2}\sinh(\nu y),
\label{Solution_1}\\
\end{equation}
The actual geometry of the problem yields the symmetric condition $v|_{y=\pm 1/2}=v_{0}$ for longitudinal velocity at the inner bar walls.
Therefore, we obtain
\begin{equation}
v(y)=1+(v_{0}-1)\frac{\cosh(\nu y)}{\cosh(\nu/2)},
\label{Solution_2}\\
\end{equation}
The Drude uniform flow follows from Eq.(\ref{Solution_2}) when $v=v_{0}=1$ and was examined
in Sec.\ref{Current induced Hall Effect in 2D bar} already. For arbitrary condition $v_{0}\neq 1$ at the inner bar walls
the profile of longitudinal velocity is expected to affected crucially by viscosity effects. The primary goal of the present paper
concerns the calculation of the sample resistivity affected by boundary conditions and viscosity strength.

A final note concerns Eq.(\ref{YX_velocity_Stokes}) which gives previous result specified by Eq.(\ref{velocity_Drude+}) for carrier drift in crossed $E_{y}\perp B_{x}$ fields when $\eta_{xy}\rightarrow 0$.

\subsection{Poiseuille viscous flow}
\label{Poiseuille viscous flow}
Let us first consider a simple case of wall adhesion condition $v_{0}=0$ \cite{Gurzhi63} followed from analysis of
Poiseuille's viscous flow known in conventional hydrodynamics. In Fig.\ref{Fig3} the blue curves depict the spatial dependence
of the flux velocity $v(y)$ specified by Eq.(\ref{Solution_2}) for different viscosity strengths. As expected, for small viscosity $\nu \gg 1$ the fluid
velocity is mostly uniform excepting ultra-narrow layer $\sim \lambda$ close to bar inner walls. In contrast, for highly viscous case $\nu \leq 1$ the flux velocity follows the Poiseuille's flow law $v(y)=\frac{\nu^{2}}{2}(\frac{1}{4}-y^{2})$ shown by the dashed line in Fig.\ref{Fig3}.

\subsection{Diamagnetic viscous flow}
\label{Diamagnetic viscous flow}
The special interest of the present paper concerns a possibility of unusual boundary condition $v_{0}>1$
whose physical background will be illustrated hereafter.

\begin{figure}[tbp]
\begin{center}\leavevmode
\includegraphics[width=0.9\linewidth]{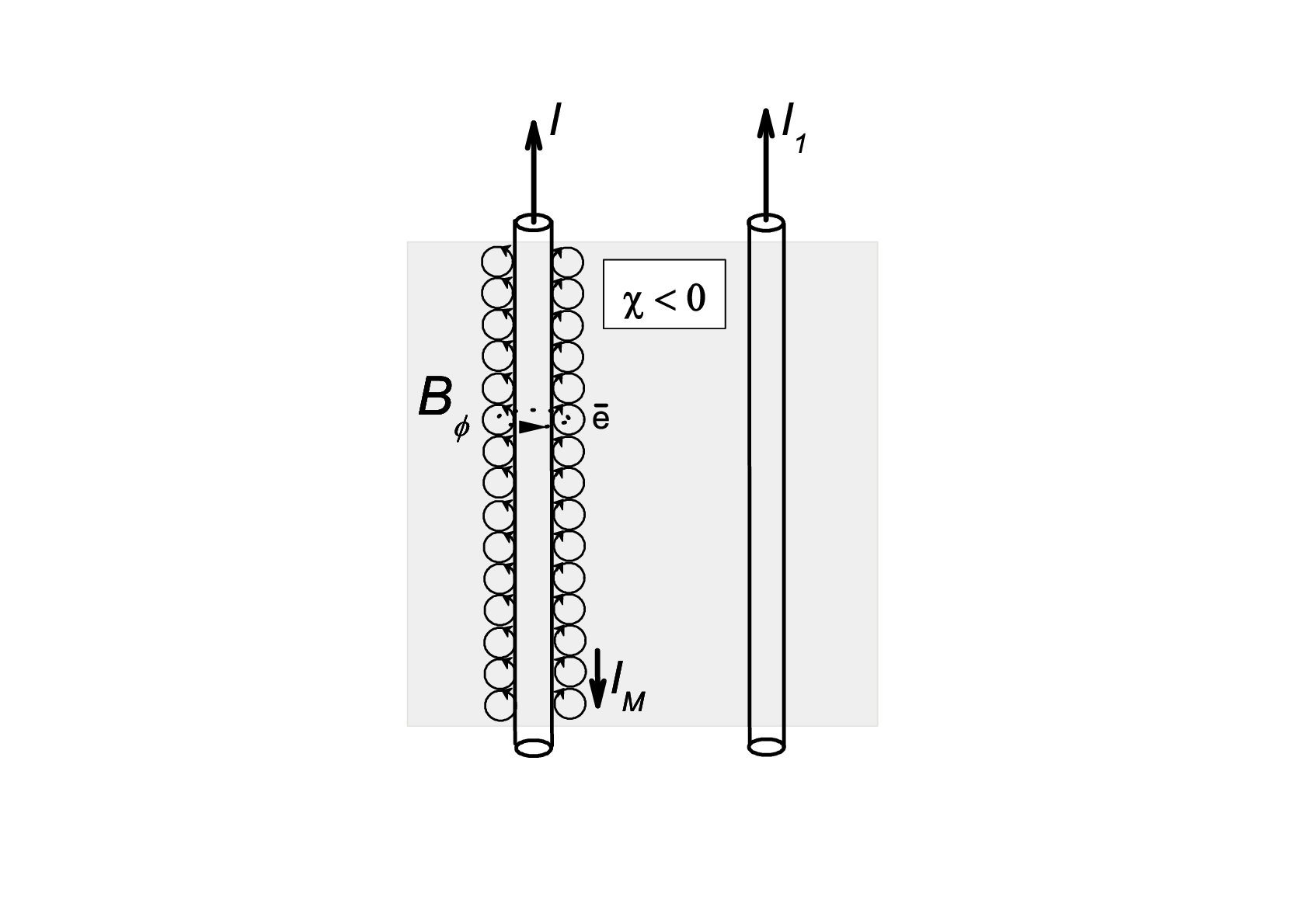} \caption[]{\label{Fig2} The macroscopic magnetic current $I_{M}$ for current carrying conductors placed into diamagnetic $\chi<0$ media (under Ref.\cite{Sivukhin96}).}
\end{center}
\end{figure}

At first, recall a scenario of a current carrying wire surrounded by diamagnetic media shown in Fig.\ref{Fig2}. The current $I$ is
provided by external source. We assume the diamagnetic susceptibility $\chi<0$ of the environment caused, for clarity, by the electrons orbital movement\cite{Landau1930}.  The current carrying wire induces the azimuthal magnetic field $B_{\varphi}=\frac{2I}{cR}$ in the surrounding space $R>R_{0}$. Notably, the magnetic field at the outer wire wall $B_{0}$ results in negative macroscopic current $I_{M}=4\pi \chi I$\cite{Sivukhin96,Vlasov05} because of diamagnetic environment. The total current flowing along the wire $I+I_{M}<I$. Let an another conductor with a co-directional current $I_{1}\parallel I$( see Fig.\ref{Fig2},a ) is placed in parallel to the initial one. Again, the total current along the second wire $(1+4\pi \chi) I_{1}$ includes the negative component $4\pi \chi I_{1}$ (not shown in Fig. \ref{Fig2},a) as well. One can check that Ampere's attractive force $\sim (1+4\pi \chi)I \cdot I_{1}$ between the pair of wires with parallel currents is reduced by a factor of $1+4\pi\chi$\cite{Sivukhin96} compared to that in absence of diamagnetic environment. Evidently, the Ampere's force diminution is caused by microscopic magnetic currents at the outer wire surface.

We now provide a strong evidence of similar effect for current carrying bar sample(see Fig.\ref{Fig1}) which exhibits diamagnetic
susceptibility $\chi<0$ itself. Indeed, for certain applied current $I$ the transverse magnetic field at internal walls of
the slab $B_{x}(\pm d/2)=\mp B_{0}$ results in the extra diamagnetic current $I_{M}$ which, in turn, is {\it collinear} to native
current, namely $I_{M}=4\pi|\chi| I$. Phenomenologically, one may imagine a diamagnetic current which flows within narrow layers of a
width $\delta$ in the vicinity of up(down) sample walls. The respective diamagnetic current density $j_{M}=\frac{I_{M}}{2\delta w}$ becomes
proportional to transverse magnetic field $B_{0}$ at the sample walls:
\begin{equation}
j_{M}=\frac{|\chi|cB_{0}}{\delta}
\label{jm_vs_B}\\
\end{equation}
and could, in principle, exceed the Drude current density $j_{D}$.

One may write down the dimensionless flux velocity $v_{0}$ at the inner rod surface as
\begin{equation}
v_{0}=\frac{j_{M}}{j_{D}}=\frac{\overline{j}}{j_{D}}\kappa,
\label{boundary_condition}\\
\end{equation}
where $\overline{j}=\frac{I}{d w}$ is the average current density. Then, we introduce the dimensionless parameter
\begin{equation}
\kappa=\frac{2\pi d |\chi|}{\delta},
\label{kappa}\\
\end{equation}
which depends on the sample size. Without diamagnetic currents, i.e. when $\kappa=0$, we recover the Poiseille's flow
when the wall-adhesion condition $v_{0}=0$ remains valid.

\begin{figure}[tbp]
\begin{center}\leavevmode
\includegraphics[width=0.9\linewidth]{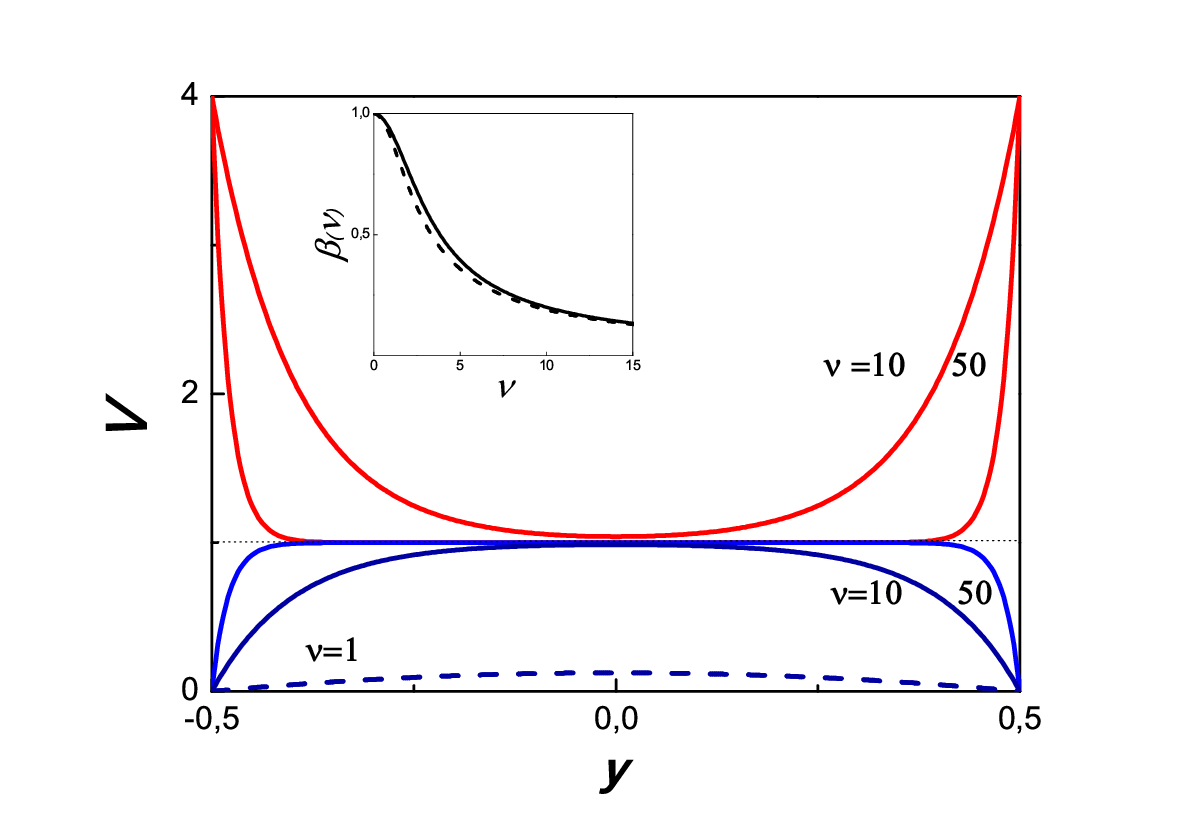} \caption[]{\label{Fig3}  Flux velocity distribution $v(y)$
specified by Eq.(\ref{Solution_2}) at fixed applied longitudinal electric field and viscosity parameter $\nu=10;50$ for
wall adhesion boundary condition $v_{0}=0$(blue) and diamagnetic boundary condition $v_{0}=4$(red).
The Poiseuille flow for $v_{0}=0;\nu=1$ is shown by the dashed line. Dotted line represents uniform flow $v=v_{0}=1$.
Inset: universal dependence $\beta(\nu)$ for bar(wire) is shown by solid(dashed) line respectively.}
\end{center}
\end{figure}

Our major interest concerns the strong diamagnetic case when $\kappa\geq 1$. In Fig.\ref{Fig3} we plot the transverse distribution $v(y)$
at fixed boundary velocity $v_{0}=4$ and different viscosity strengths. As expected, the diamagnetic current within a narrow layer $\delta$ initiates a current flow within a wider stripe $\lambda \gg \delta$ close to sample inner wall. The flux velocity approaches the Drude value in a sample bulk, i.e. when $v=1$. Using Eq.(\ref{Solution_2}) we find out the average current density $\overline{j}=\frac{ne}{d}\int\limits^{d/2}_{-d/2}V_{z}(Y) dY$ as it follows
\begin{equation}
\overline{j}=j_{D}\left [ 1+(v_{0}-1)\beta(\nu)\right ],
\label{Average_Curent}\\
\end{equation}
where $\beta(\nu)=\frac{2}{\nu}\tanh (\frac{\nu}{2})$ is the universal function(see Fig.\ref{Fig3},inset) of the viscosity strength.
The function $0<\beta(\nu)\leq 1$ decreases smoothly as $\sim 1-\nu^{2}/12$ for high-viscous case $\nu \ll 1$ and, then follows
the asymptote $\sim 2/\nu$ for low viscosities $\nu \gg 1$.

Remarkably, the all previous reasoning are valid for a wire whose radius plays the role of the sample thickness $d$ in present notations.
For wire the universal function $\beta(\nu)$ embedded into Eq.(\ref{Average_Curent}) can be replaced by $\beta_{wire}(\nu)=\frac{2J_{1}(\nu)}{\nu J_{0}(\nu)}$, where $J_{0(1)}$ is the zero(first)-order modified Bessel function of the first kind. The dependence $\beta_{wire}(\nu)$ is shown by the dashed line in Fig.\ref{Fig3},inset. Both dependencies are close one to each other, therefore our forthcoming discussion could be similar for wire case as well.

Using Eq.(\ref{boundary_condition}) the self-consistent solution of Eq.(\ref{Average_Curent}) reads
\begin{equation}
\overline{j}=j_{D}\frac{1-\beta}{1-\kappa \beta}.
\label{SC_current}\\
\end{equation}
Noticeably, Eq.(\ref{SC_current}) defines the average current density at fixed longitudinal electric field $E_{z}$.
Consequently, one may define the "effective resistivity" $\rho=\frac{E_{z}}{\overline{j}}$ as it follows
\begin{equation}
\rho=\rho_{D}\frac{1-\kappa \beta}{1-\beta}.
\label{SC_resistivity}\\
\end{equation}
Here, $\rho_{D}=\frac{m}{ne^{2}\tau}$ is the conventional Drude resistivity, $\mu_{D}=(ne\rho_{D})^{-1}$ is the mobility.

\begin{figure}[tbp]
\begin{center}\leavevmode
\includegraphics[width=0.9\linewidth]{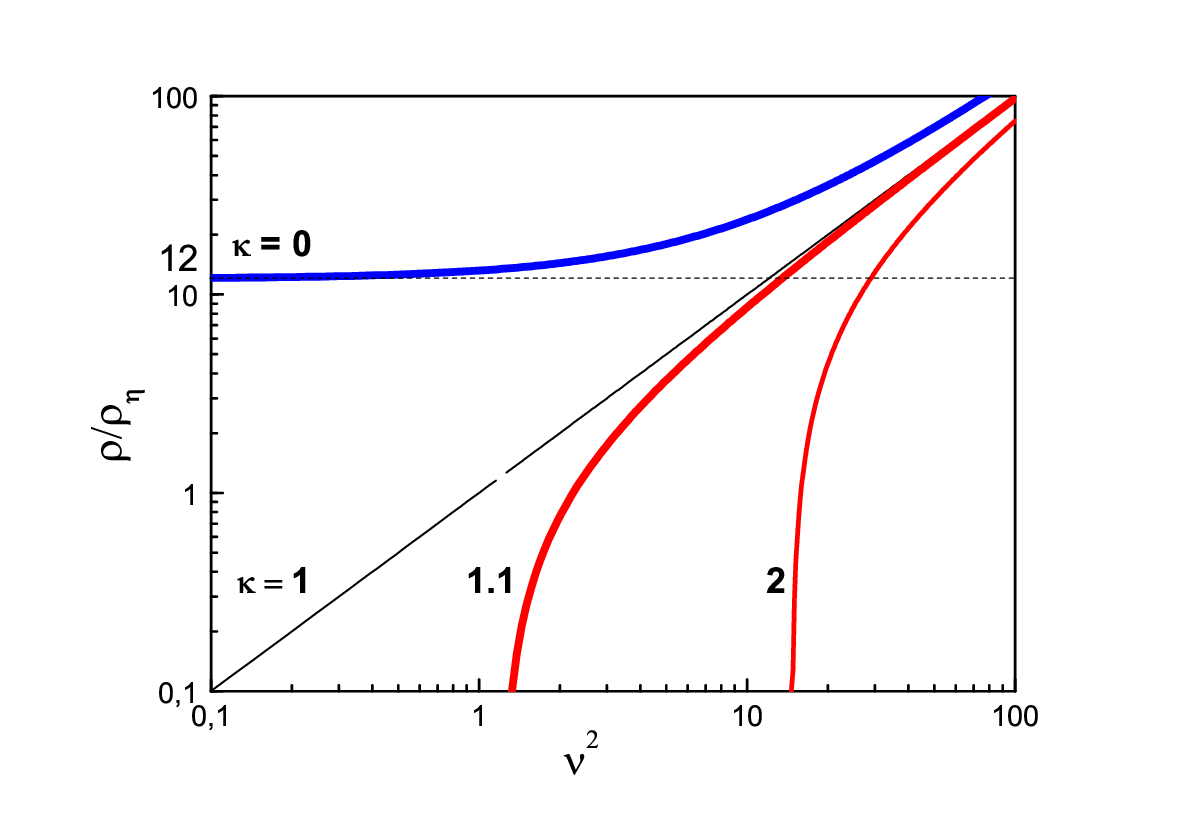} \caption[]{\label{Fig4} Dimensionless resistivity $\rho/\rho_{\eta}$ followed from
Eq.(\ref{SC_resistivity}) vs dimensionless disorder strength $\nu^{2}=\frac{d^{2}}{\eta \tau}$ for: zero
diamagnetic current $\kappa=0$; uniform current state $\kappa=1$; strong diamagnetism $\kappa>1$. Dashed line
represents the viscous resistivity $\rho=12\rho_{\eta}$ at $\kappa=0$ and $1/\tau \rightarrow 0$.}
\end{center}
\end{figure}

Eq.(\ref{SC_resistivity}) represents the central result of the paper. The galvanic measurements provide the "effective
resistivity" which differs with respect to Drude value. Namely, the "effective resistivity" depends on the size and,
moreover, the diamagnetic properties of the sample. At first, for $\kappa=1$ one recover the Drude uniform current flow without viscous
effects included, hence $\rho=\rho_{D}$. Secondly, the wall adhesion condition $\kappa=0$ provides the "effective
resistivity" in a Gurzhi form $\rho=\rho_{D}/(1-\beta)$ reported in Ref.\cite{Gurzhi63}. For low-viscous case $\nu \gg 1$
the "effective resistivity" is still described by Drude formulae $\rho \sim \rho_{D}$. In the opposite high-viscosity
and(or) low dissipation limit $\nu \ll 1$ the Poiseille type of a current flow is realized. The "effective resistivity" at $\nu \ll 1$
is given by the asymptote $\rho=12\rho_{\eta}$, where $\rho_{\eta}=\frac{m}{ne^{2}}\frac{\eta}{d^{2}}$ is
so-called "viscous" resistivity\cite{Gurzhi63} which depends on the sample size.  Note, the ratio $d^{2}/\eta$ plays the role
of the momentum relaxation time similar to that discussed\cite{Alekseev16,Shi14} for 2D electron gas. The transition from Drude to "viscous"
resistivity case occurs at $\nu\sim 1$. In Fig.\ref{Fig4} we plot the reduced resistivity $\rho/\rho_{\eta}$ vs
disorder $\nu^{2}\sim 1/\tau$ for fixed viscosity strength $\eta$ and different valued of diamagnetic parameter $\kappa$.
Noticeably, for arbitrary value of the diamagnetic parameter $\kappa$ the resistivity in Fig.\ref{Fig4} starts to follow conventional
Drude dependence for high disorder and(or) low viscosity $\nu \gg 1$.

The most intriguing result followed from Eq.(\ref{SC_resistivity}) concerns the strong diamagnetism case $\kappa>1$ when the
effective resistivity vanishes at
\begin{equation}
\kappa \cdot \beta(\nu)=1.
\label{Critical_condition}\\
\end{equation}
Eq.(\ref{Critical_condition}) gives the critical condition for zero resistivity, i.e. the superconductivity\cite{Kamerlingh-Onnes1911}.
Recall that for arbitrary argument $\beta(\nu) \leq 1$. Hence, the superconductivity would appear when the condition $\kappa>1$ is satisfied. The latter
can be re-written in terms of film size as $d \geq d_{m}$, where we make use of minimal sample thickness
\begin{equation}
d_{m}=\frac{\delta}{2\pi |\chi|},
\label{Minimal_wire_radius}\\
\end{equation}
for which the superconductivity can be realized. We further demonstrate that superconductivity criteria $\kappa=d/d_{m}>1$ is even stronger
for real systems.

We emphasize that the superconductivity may appear for even finite momentum relaxation time. This result looks like mysterious
at a first glance. Nevertheless, the experimental data\cite{Meissner32} provide a strong evidence of
the disorder remains finite whenever the superconductivity is present. A finite momentum relaxation time
was estimated\cite{Meissner32} for lead as $T \rightarrow 0$.

The actual physics of superconductivity is rather transparent. The non-dissipative diamagnetic current is pinched within
a narrow inner layer $\lambda \sim d_{m}/2$ of a slab and, hence shunts the dissipative Drude flow in the sample bulk.
The total current in a sample becomes purely diamagnetic when Eq.(\ref{Critical_condition}) is fulfilled.

\subsection{Size effect of superconductivity transition}
\label{Size effect of superconductivity transition}

\begin{figure}[tbp]
\begin{center}\leavevmode
\includegraphics[width=1.0\linewidth]{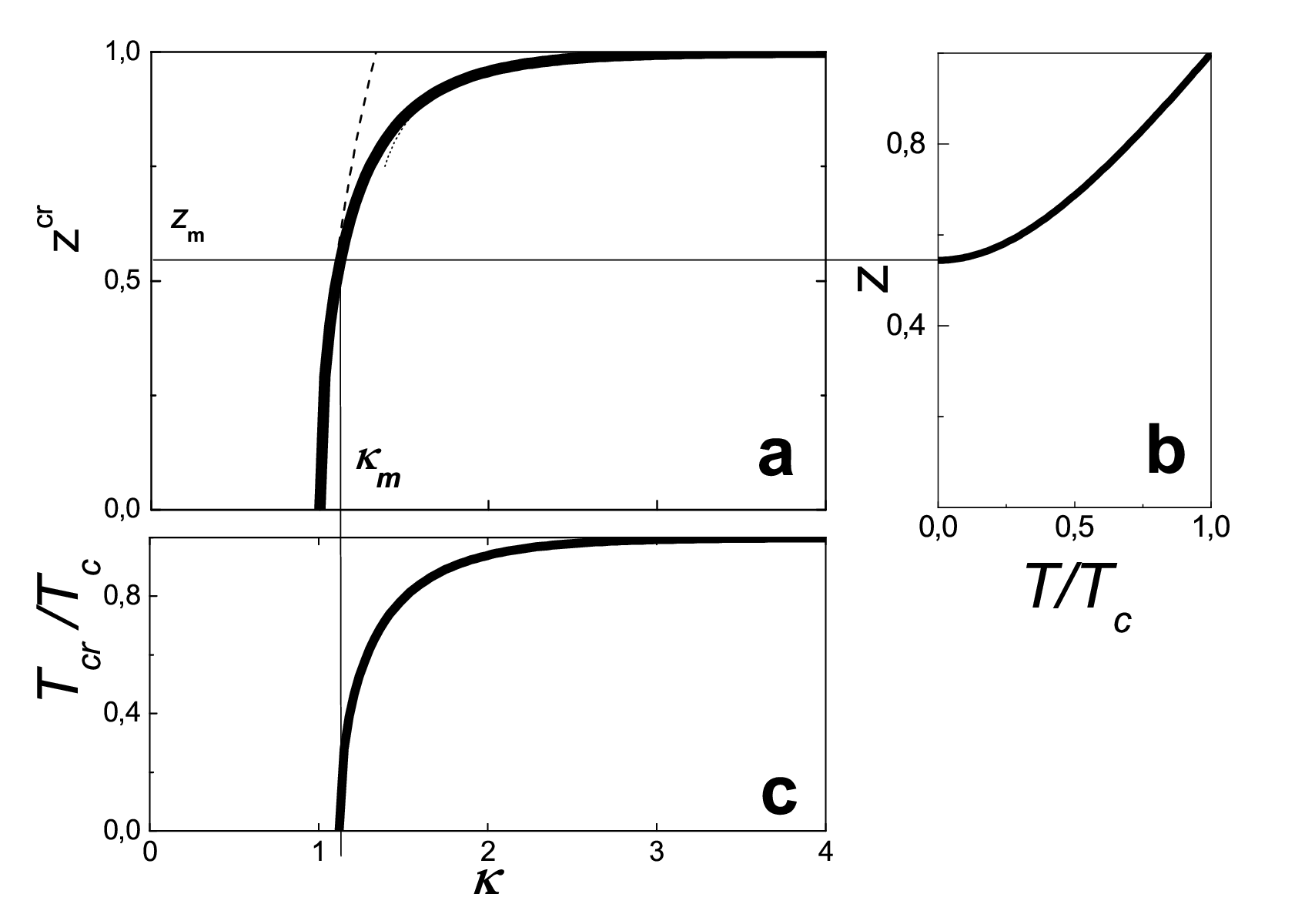} \caption[]{\label{Fig5} a) The critical diagram $z^{cr}(\kappa)$ of superconductivity followed from Eq.(\ref{transcendental_equation}). The asymptotes for small $\kappa-1\ll 1$ and bulky sample $\kappa \gg 1$ are shown by dashed and dotted line respectively. The area below the critical curve $z^{cr}(\kappa)$ corresponds to superconductivity. b) The dependence $z(\Theta)$ specified by Eq.(\ref{z(T)plus}); c) Temperature threshold $\Theta_{cr}(\kappa)$ specified by Eq.(\ref{Threshold_temperature}) for fixed $z_{m}=0.54$.}
\end{center}
\end{figure}

We now examine in greater details the critical condition given by Eq.(\ref{Critical_condition}). One can find, in principle,
the critical dependence in a form $\nu^{cr}(\kappa)$. The latter is, however, non-informative since both
variables $\kappa,\nu$ depend on the sample size. To avoid this problem, let us introduce a size-free parameter
$z=\frac{\nu}{2\kappa}=\frac{d_{m}}{2\lambda}$. With the help of the above notation
the modified Eq.(\ref{Critical_condition}) yields
\begin{equation}
\kappa=\frac{\text{arctanh}(z)}{z}
\label{transcendental_equation}\\
\end{equation}
and denotes the desired critical diagram in a compact form since $z^{cr}(\kappa)<1$. The latter is shown in Fig.\ref{Fig5},a. The area below the critical curve
corresponds to superconductivity. For sample thickness closed to its minimal value $d_{m}$, i.e. when $\kappa-1\ll 1$, the critical curve follows the asymptote $z=\sqrt{3(\kappa-1)}$ depicted by the dashed line in Fig.\ref{Fig5},a. Then, the critical curve saturates asymptotically as $\kappa=\ln(\frac{2}{1-z})/(2z)$ for bulky sample, i.e. when $\kappa\gg 1$.

We now find superconductivity threshold in terms of temperature by looking at parameter $z\sim 1/\lambda=1/\sqrt{\eta\tau}$. Remind that
for actual low-temperature case the transport is governed mostly by scattering on static defects, hence one may consider a constant momentum relaxation time
$\tau \neq \tau(T)$. A point of interest is the scattering time $\tau_{\eta}$ embedded into components of viscosity tensor specified by Eq.(\ref{viscosity_components}). A common belief is that the carrier viscosity is caused solely by e-e collisions with the reciprocal scattering time\cite{Pomeranchuk36,Quinn58,Abrikosov59}
\begin{equation}
\frac{1}{\tau_{ee}(\xi)}=\frac{\xi^{2}}{\tau_{1}}
\label{ee_time}\\
\end{equation}
dependent on the reduced temperature $\xi=T/T_{F}$. Here, $T_{F}=\varepsilon_{F}/k$ and $\varepsilon_{F}$ are the Fermi temperature and energy
respectively. Then, $\tau_{1}=l_{1}/V_{F} \sim  \hbar/\varepsilon_{F}$\cite{Quinn58} is the path time of interelectronic distance $l_{1}\sim n^{-1/3}$. At low temperatures the e-e collisions becomes strongly dumped because of Pauli principle, therefore the e-e mean free path $l_{ee}=\tau_{ee}V_{F}$ becomes infinite. Consequently, one expects an infinite electron viscosity. The whole story is that apart from e-e scattering contribution any process yielding the
relaxation of the second moment of the electron distribution function(for example, the scattering on static defects) would
influence\cite{Steinberg58,Alekseev16} the electron viscosity. With the aid of standard Mattissen's rule, one may write down the reciprocal viscosity length
as it follows
\begin{equation}
\frac{1}{l_{\eta}}=\frac{\xi^{2}}{l_{1}}+\frac{1}{l_{0}},
\label{visc_relax_time}\\
\end{equation}
where $l_{0}$ defines the residual value at $T \rightarrow 0$. The hydrodynamic approach is valid when $l_{\eta} \leq l$. Further, we will use recent
argumentation\cite{Cheremisin21} and verify the above criterion for actual case of dirty metals.

With the help of Eq.(\ref{visc_relax_time}) the variable $z=\frac{d_{m}}{2\lambda}$ becomes temperature dependent, namely
\begin{equation}
z(\xi)= z_{m} \sqrt{1+\gamma \xi^{2}},
\label{z(T)}\\
\end{equation}
where $z(0)=z_{m}=\frac{\sqrt{5}d_{m}}{2\sqrt{l_{0}l}}$ is the zero-temperature value, $\gamma=l_{0}/l_{1}$ is the dimensionless ratio. Recall that threshold diagram in Fig.\ref{Fig5}a demonstrates $z^{cr}(\kappa)\leq 1$. Hence, the condition $z_{m} \leq 1$ must be satisfied for superconductivity to be feasible. We re-write this criteria in terms of the carrier mobility as it follows
\begin{equation}
\mu\geq\mu_{\text{min}},
\label{condition_Zm}\\
\end{equation}
where $\mu_{\text{min}}= \frac{5}{8}\frac{ed_{m}^{2}}{\varepsilon_{F}\tau_{0}}$ plays the role of the minimal
mobility for which superconductivity is possible. Further, we will use obvious relationship
\begin{equation}
z_{m}=\sqrt{\mu_{\text{min}}/\mu}
\label{definition_Zm}\\
\end{equation}
as well.

If $z_{m}<1$, the only upper part of the threshold diagram in Fig.\ref{Fig5},a remains useful, i.e. when $z^{cr}(\kappa)>z_{m}$.
Then, the equality $z_{m}=z^{cr}(\kappa_{m})$ denotes a certain value
of minimal sample size parameter $\kappa_{m}$, which corresponds to superconductivity threshold at $T=0$. Evidence shows that at finite
temperature the superconductivity can be realized for samples whose sizes satisfy the condition $\kappa\geq\kappa_{m}$.
The latter gives the strict criteria for minimal sample size
\begin{equation}
d\geq d_{m}\cdot \kappa_{m}
\label{minimal_sample_size}\\
\end{equation}
instead of that $d\geq d_{m}$ discussed earlier.

We now attempt to find out threshold temperature for the most important case of massive sample $\kappa \gg 1$ known to be a
universal quantity\cite{Kamerlingh-Onnes1911}. Indeed, substituting the condition $z(\xi)=1$ valid for massive sample into
Eq.(\ref{z(T)}) we find out the sought-for threshold temperature
\begin{equation}
T_{c}= T_{F} \left[ (\mu/\mu_{\text{min}}-1)/\gamma \right]^{1/2}.
\label{T_c}\\
\end{equation}
For clarity, we will label hereafter the all quantities related to superconductivity threshold for bulky sample by index "c".
According to Eq.(\ref{T_c}), the superconductivity is possible for massive specimen when $\mu\geq\mu_{{\text min}}$. The
better the sample quality the higher the threshold temperature.

We emphasize that for finite size sample the threshold temperature is always less than that for bulky samples. Experimentally,
the drop in threshold temperature for smaller samples was first reported in Refs.\cite{Meissner33,Meissner33+}. It is instructive to use the dimensionless
temperature $\Theta=T/T_{c}$ scaled with respect to threshold temperature of bulky sample. Hence, the modified Eq.(\ref{z(T)}) yields
\begin{equation}
z(\Theta)= \sqrt{z_{m}^{2}+(1-z_{m}^{2})\Theta^{2}},
\label{z(T)plus}\\
\end{equation}
In Fig.\ref{Fig5},b we plot the dependence given by Eq.(\ref{z(T)plus}). Combining the dependencies $z(\Theta)$ and $z^{cr}(\kappa)$
specified by Eq.(\ref{z(T)plus}) and Eq.(\ref{transcendental_equation}) respectively we obtain the reduced threshold
temperature $\Theta_{cr}=T_{cr}/T_{c}$ as a function of the sample size
\begin{equation}
\Theta_{cr}(\kappa) = \left[ \frac{z^{cr}(\kappa)^{2}-z^{2}_{m}}{1-z^{2}_{m}} \right]^{1/2}.
\label{Threshold_temperature}\\
\end{equation}
Fig.\ref{Fig5},c demonstrates an example of threshold dependence given by Eq.(\ref{Threshold_temperature}). As expected, $T_{cr}=T_{c}$ for massive sample.

\begin{figure}[tbp]
\begin{center}\leavevmode
\includegraphics[width=1.4\linewidth]{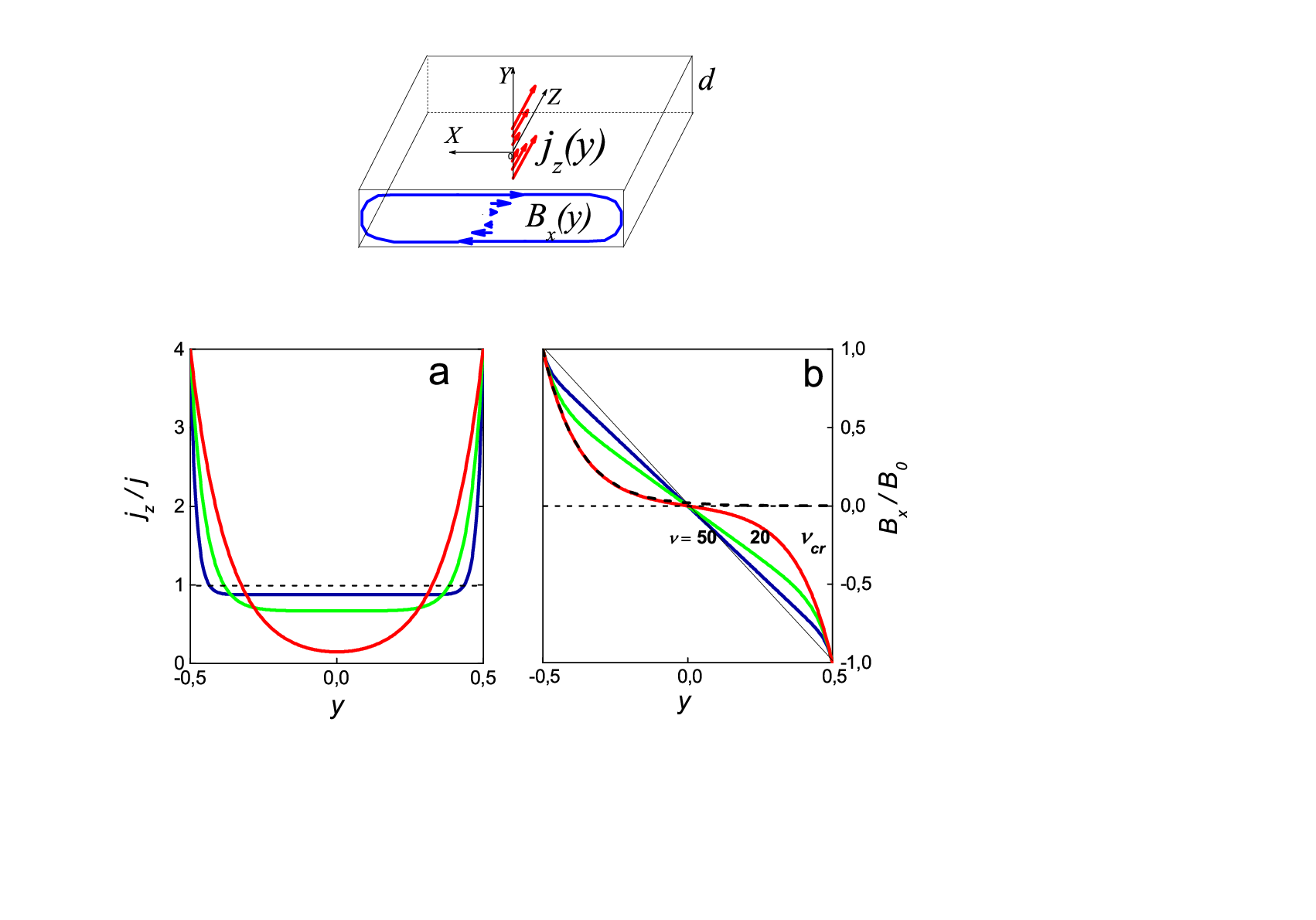} \caption[]{\label{Fig6} Top panel: Experimental setup. Dimensionless current density $j_{z}/j_{M}$(panel a) and azimuthal magnetic field $B_{x}/B_{0}$ (panel b) specified by Eq.(\ref{J_B real}) for finite size sample at $\kappa=4$(corresponds to $z^{cr}=0.999$ and
$\nu_{cr}=2\kappa z^{cr}\cong 8$) and viscosity parameter $8;20;50$. Thin lines depict the uniform current density case when $\kappa=1$. Dotted line(insert) corresponds to magnetic field screening asymptote described in text.}
\end{center}
\end{figure}

\subsection{Magnetic field screening}
\label{Magnetic field screening}
We now find both the current density and magnetic field spatial profiles for system closer to superconductivity threshold.
Recall that the all previous discussion concerned the presence of a finite electric field $E_{z}$. Aiming
to account for superconductivity(i.e. when $E_{z}=0$) we inverse Eq.(\ref{SC_current}) and, then find the electric field as a
function of the average current density $\overline{j}$. Combining the result with Eq.(\ref{Solution_2}) we obtain the flux velocity
distribution in terms of the average current density. The current density profile $j_{z}(y)$ and transverse
magnetic field $B_{x}(y)$ specified by Eq.(\ref{Transversed_Field}) read:
\begin{eqnarray}
j_{z}(y)=\overline{j} \left [ \frac{1-\kappa \beta}{1-\beta}-\frac{1-\kappa}{1-\beta} \cdot \frac{\cosh(\nu y)}{\cosh(\nu/2)} \right ],
\eqnum{1} \label{J_B real}\\
B_{x}(y)=B_{0}\left [-2\frac{1-\kappa \beta}{1-\beta}y+\beta \cdot \frac{1-\kappa}{1-\beta}\cdot \frac{\sinh(\nu y)}{\sinh(\nu/2)} \right ].
\nonumber
\end{eqnarray}
Eq.(\ref{J_B real}) gives the correct values of the diamagnetic current density $j_{M}$ and the amplitude of the magnetic field $B_{0}=\frac{2\pi}{c}\overline{j}d$ at the inner walls of the sample $y=\pm 1/2$. For $\kappa=1$ one recovers the result for uniform current flow, namely $j_{z}(y)=\overline{j}$, $B_{x}=-2B_{0}y$ . As an example, the dependencies given by Eqs.(\ref{J_B real}) are plotted in Fig.\ref{Fig6} for fixed diamagnetic parameter $\kappa=4$. Evidence shows that fluid viscosity enhancement leads to progressive shift of the current towards the inner walls of the bar. Simultaneously, the magnetic field is pushed out from the sample bulk.

It is of particular interest the current density and(or) magnetic field profiles for superconductivity. Combining Eq.(\ref{Critical_condition}) and Eqs.(\ref{J_B real}) one obtains
\begin{eqnarray}
j_{z}(y)=\overline{j}\cdot \kappa \frac{\cosh(\nu y)}{\cosh(\nu/2)},
\label{J_ZRS}\\
B_{x}(y)=-B_{0} \frac{\sinh(\nu y)}{\sinh(\nu/2)}.
\label{B_ZRS}\\
\nonumber
\end{eqnarray}
Note that Eq.(\ref{J_ZRS}) follows immediately from Navier-Stocks Eq.(\ref{XX_velocity_Stokes}) when one puts $E_{z}=0$. One may perform the integration
of Eq.(\ref{J_ZRS}) using Maxwell equation $\text{rot}\textbf{B}=\frac{4\pi}{c}\textbf{j}$. In such a way we recover Eq.(\ref{B_ZRS})
and, moreover,confirm again the threshold condition $\kappa\beta=1$.

\begin{figure}[tbp]
\begin{center}\leavevmode
\includegraphics[width=1.0\linewidth]{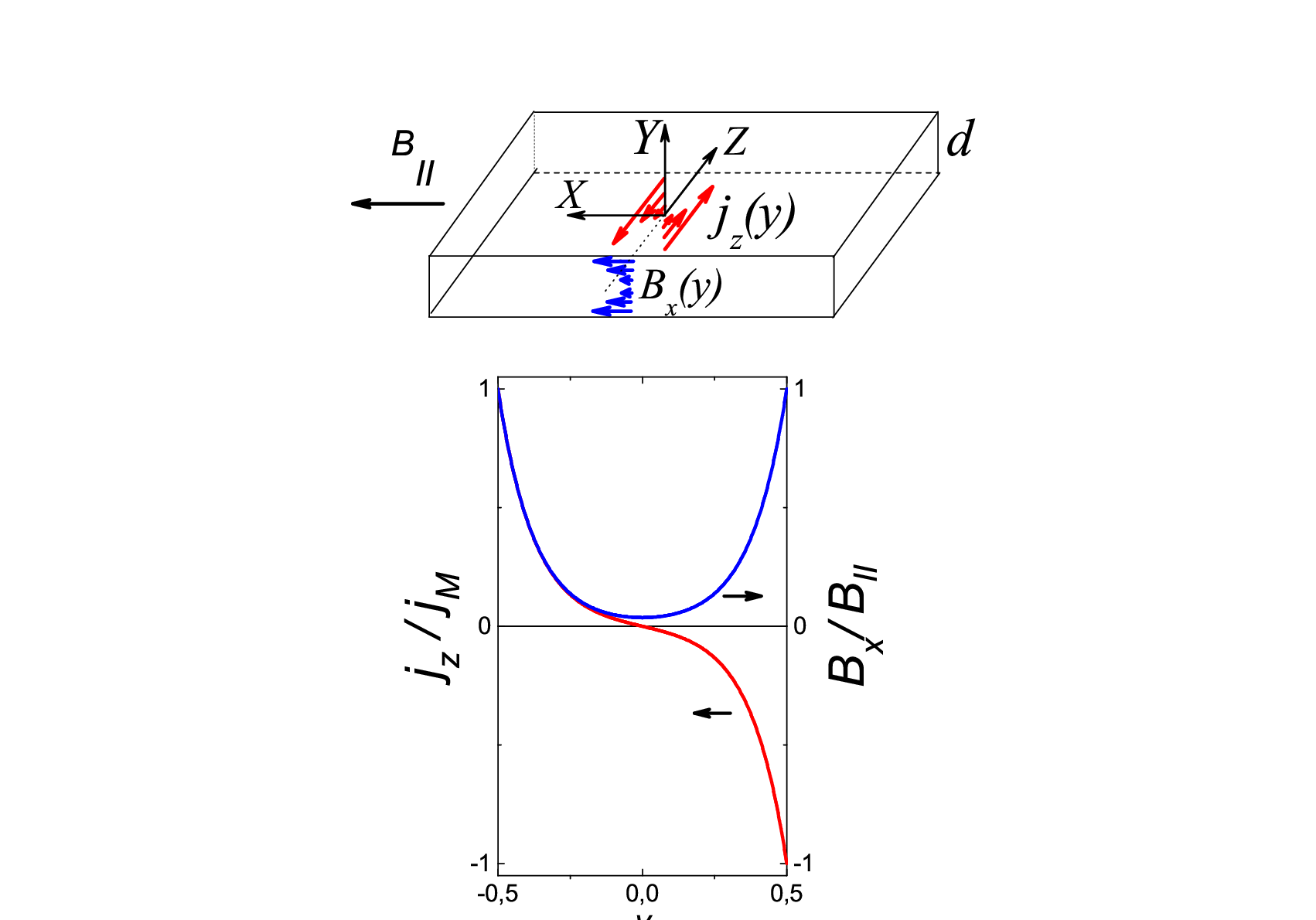} \caption[]{\label{Fig7} Experimental setup (top panel) for thin sample placed in parallel
field. Bottom panel: dimensionless current density $j_{z}/j_{M}$ and azimuthal field $B_{x}/B_{0}$ profiles specified by
Eqs.(\ref{B(B Fixed)},\ref{J(B Fixed)}) for superconductivity at $\kappa=4$ and $\nu_{cr}=8$.}
\end{center}
\end{figure}

Recall that the current(field) penetration length $\lambda=\sqrt{\eta\tau}=\frac{d_{m}}{2z(\Theta)}$ depends on the temperature.
At high temperatures the fluid is non-viscous therefore the penetration length is small since $z\rightarrow \infty$. At low temperatures the penetration
length increases. For bulky specimen at threshold temperature $T=T_{c}$ the penetration length is given $\lambda=\frac{d_{m}}{2}$.
To confirm this finding, we put in Fig.\ref{Fig6},b the exponential asymptote $B_{x}=B_{0}\exp(-2\kappa(y+1/2))$ imposed to
magnetic field profile specified by Eq.(\ref{B_ZRS}).

It is of considerable interest the case of a specimen placed into external magnetic field $B_{\parallel}$ parallel to the
slab plane. Evidence shows that current distribution depicted in Fig.\ref{Fig7} may
prevent magnetic field penetration to the sample bulk. Again, for $E_{z}=0$ the Navier-Stocks Eq.(\ref{XX_velocity_Stokes}) allows one to find
the current density profile
\begin{equation}
j_{z}(y)=-j_{M} \frac{\sinh(\nu y)}{\sinh(\nu/2)}.
\label{J(B Fixed)}\\
\end{equation}
After subsequent integration one obtains the magnetic field distribution in the sample bulk
\begin{equation}
B_{x}(y)=B_{\parallel} \frac{\cosh(\nu y)}{\cosh(\nu/2)}.
\label{B(B Fixed)}\\
\end{equation}
where $B_{\parallel}=\frac{4\pi}{c} j_{M} \frac{d}{\nu} \tanh^{-1}(\nu/2)$. For actual geometry Eq.(\ref{jm_vs_B}) reads $j_{M}=\frac{cB_{\parallel}}{2\pi d}\kappa$, therefore we obtains a modified threshold condition $\kappa\beta/\tanh^{2}(\nu/2)=1$ for superconductivity. Intriguingly, for bulky sample $\nu \rightarrow \infty$ the latter coincides with threshold condition given by Eq.(\ref{Critical_condition}) for current carrying sample which confirm, in general,
the Silsbee's\cite{Silsbee1916} hypothesis.

\section{Transition from normal state to superconductivity}
\label{Transition from normal state to superconductivity}

\subsection{Sample-size driven normal state to superconductivity transition}
\label{Sample-size driven normal state to superconductivity transition}

\begin{figure}[tbp]
\begin{center}\leavevmode
\includegraphics[width=1.0\linewidth]{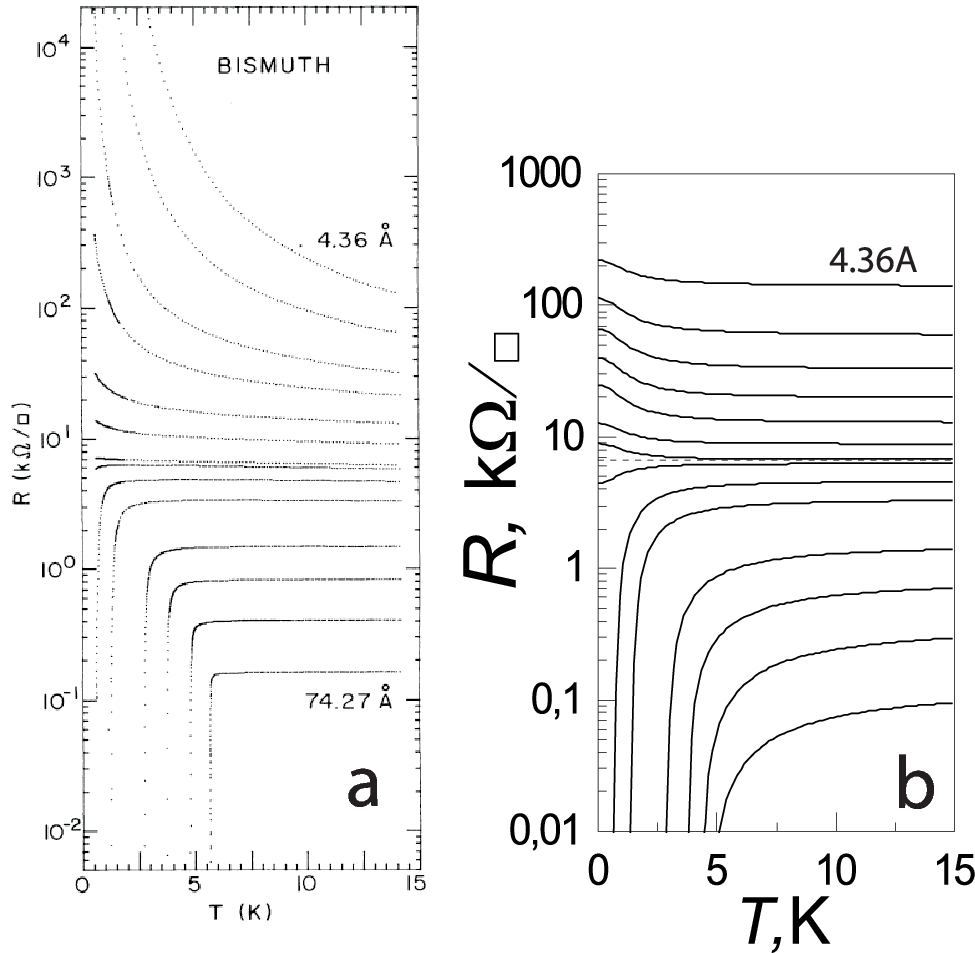} \caption[]{\label{Fig8} Sheet resistance $R_{\square}$ vs temperature for amorphous bismuth thin films: a) experimental data under Ref.\cite{Haviland89} for sample thickness $d[A]$=63.7;25;14.8;10.7;7.9;7.4;6.7;6.4;6.2;5.7;5.4;5.0;4.65,4.36. b) calculated.}
\end{center}
\end{figure}

In Sec.\ref{Size effect of superconductivity transition} we have demonstrated a possibility of the superconductivity phase
for film thickness $d\geq d_{m}\kappa_{m}$. For thinner samples the resistivity remains finite. One expect that the shape of
the temperature dependence of the resistivity would be sensitive to variation of the sample thickness. This effect is known in
literature\cite{Jaeger89,Haviland89,Liu93} as so-called sample-size driven normal-to-superconductor state transition. Remarkably,
our model predicts such a transition which has been shown in Fig.\ref{Fig4} already.
Indeed, for sample thickness changed from bigger $\kappa>1$ to a smaller $\kappa<1$ one, the shape of the resistivity curves $\rho(\nu)$ in
Fig.\ref{Fig4} changes drastically. Obviously, the temperature dependence of the resistivity duplicates this tendency since $\nu\sim 1/\sqrt{\eta}\sim T $.
Actually, one would expect an abrupt change from "metallic" upturn $\frac{d\rho}{dT}>0$ to "insulating" downturn $\frac{d\rho}{dT}<0$
behavior as the sample size decreases. For $\kappa=1$ the resistivity is constant given by Drude value $\rho=\rho_{D}$. Evidence shows
that the experimental data for quench-condensed amorphous bismuth\cite{Haviland89} reproduced in Fig.\ref{Fig8}a would be an example
which supports our model predictions. We now intend to account for this effect. By the primary step, we found the correct thickness $d=63.7$A
for certain sample instead of that $d=74.27$A\cite{Haviland89,Liu93} stated in Fig.\ref{Fig8}a.

Let us discuss first the properties of amorphous bismuth whose structure was analyzed in details in Ref.\cite{Komnik73}.
It was confirmed that the molecular beam condensation of bismuth at helium temperature\cite{Buckel54} results in the
uniform filling of the substrate with atoms whose diffusion mobility is negligibly small. The film structure
remains stable being characterized by a few weakly pronounced coordination spheres. Thicker layers $\geq 100$A have a definitive
structure with the first coordination sphere of radius $\sim 3.2 $A and 4-5 atoms included. Over the temperature range explored in
Ref.\cite{Haviland89} the layers of thicknesses, down to monatomic, exhibit the metallic conductivity. According to Ref.\cite{Komnik73},
the conductivity increases for less disordered layers of enhanced thickness. In view of immutable position of
add-on Bi atoms on the film the carrier density is conventionally\cite{Komnik73} believed to be a constant governed by
valence electrons. However, the density of the conducting electrons was reliably measured\cite{Hunderi75} for only thick
films $\geq 150$A, which is much greater than those explored in Ref.\cite{Haviland89}. We argue that conventional
Hall measurements are highly desirable for reliable determination of carrier density.

We draw attention to the existence of a T-independent plane separatrix of the resistance data in Figure 8a. The latter corresponds to sample of thickness $d\simeq 6.5$A and the sheet resistance $R_{\square}\simeq 6.5$kOhm counted for normal state at $T=14$K. The separatrix matches the condition $\kappa=1$ in our notations. Thus, we put $d_{m}=6.5$A. Noticeably, the bulk resistivity $\rho_{D}=R_{\square}d_{m}=420\mu\Omega \text{cm}$ is comparable to typical value $\sim 200\mu\Omega \text{cm}$ known from Ioffe-Regel\cite{Ioffe60} criteria $l\sim a$ for dirty metal, where $a$ is the interatomic distance. Further, we will discuss
the problem of transport in dirty metals. At a moment, we concentrate on the regular conductivity $\sigma=\rho^{-1}$ being, as usual, a measure
of the disorder strength.

The careful analysis of the data in Fig.\ref{Fig8}a reveals the evidence of the superconductor transition at $T_{cr}=0$ expected for sample of thickness $d\simeq 7.2$A and the normal sheet resistance $R_{\square}\simeq 6$kOhm at $T=14$K. The respective bulk resistivity yields $\rho_{0}=R_{\square}d=410\mu\Omega \text{cm}$ at $T=14$K. This result is of extreme interest since it allows to deduce the key parameters of our model. Indeed, one may find the reduced sample size $\kappa_{m}=d/d_{m}=1.12$ and, in turn, parameter $z_{m0}=0.54$ from the critical plot in Fig.\ref{Fig5}a.
\begin{figure}[tbp]
\begin{center}\leavevmode
\includegraphics[width=1.0\linewidth]{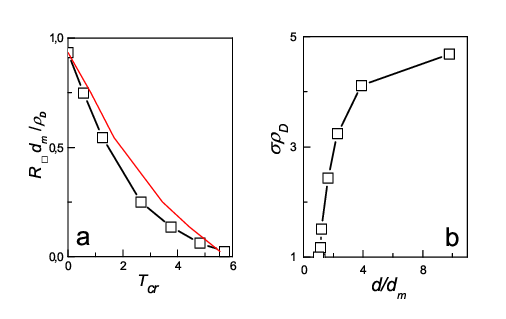} \caption[]{\label{Fig8a} a.Reduced sheet resistance $R_{\square}d_{m}/\rho_{D}$ at $T=14$K vs critical temperature of superconductivity $T_{cr}$. Theoretical result is shown by the red curve. b.Dimensionless conductivity $\sigma\rho_{D}$ vs reduced
sample thickness $\kappa=d/d_{m}$. The values $\rho_{D},d_{m}$ correspond to separatrix curve in Fig.\ref{Fig8}a.}
\end{center}
\end{figure}

\begin{figure}[tbp]
\begin{center}\leavevmode
\includegraphics[width=1.0\linewidth]{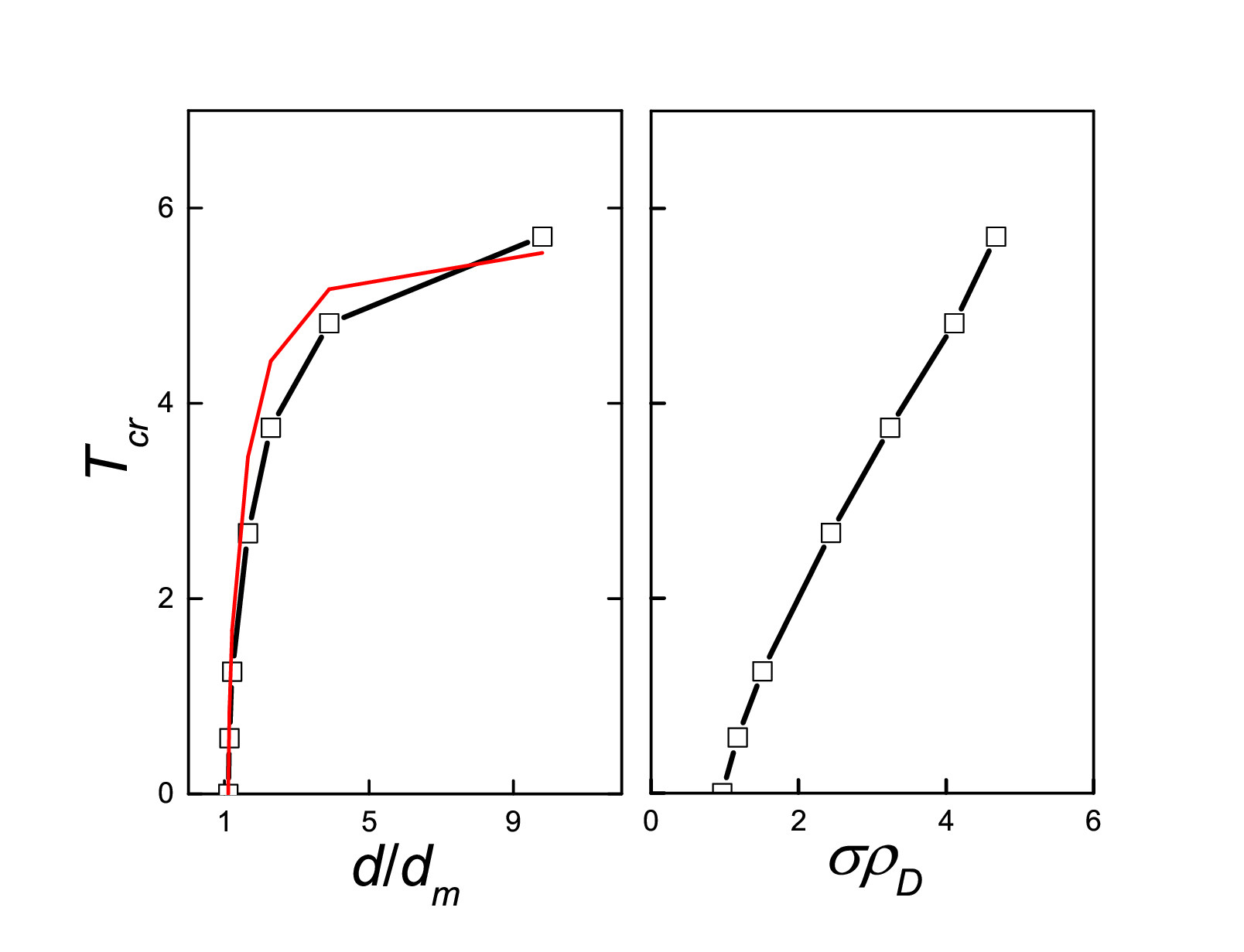} \caption[]{\label{Fig8b} Critical temperature $T_{cr}$ deduced from Fig.\ref{Fig8}a vs: a) sample size $\kappa=d/d_{m}$ and b) reduced conductivity $\sigma\rho_{D}$. The red curve shows the best fit with the help of Eq.(\ref{T_cr}).}
\end{center}
\end{figure}

Recall that for bulky amorphous bismuth the critical temperature is known $T_{c}=6.1$K. Noticeably, at elevated temperatures the all resistance curves below separatrix in Fig.\ref{Fig8}a demonstrate a flat behavior indicating the predominant role of structure assisted disorder\cite{Komnik73}. For each curve below the separatrix we use the data\cite{Haviland89} to extract the critical temperature $T_{cr}<T_{c}$, reduced layer thickness $\kappa$ and, then the transport coefficients: $R_{\square}$, $\rho=R_{\square}d$, $\sigma=1/\rho$ at $T=14$K. Following the standard practice, we plot in Fig.\ref{Fig8a}a the sheet resistance vs critical temperature $R_{\square}(T_{cr})$. Moreover, we plot the dependence of conductivity vs sample thickness $\sigma(\kappa)$ in Fig.\ref{Fig8a}b. As in Ref.\cite{Komnik73} the disorder becomes stronger in thinner films. Combining the dependencies $R_{\square}(T_{cr})$ and $\sigma(\kappa)$ we obtain the temperature threshold either in terms of reduced size $T_{cr}(\kappa)$ or conductivity $T_{cr}(\sigma)$. Both dependencies are illustrated in Fig.\ref{Fig8b}. We first focus on the critical diagram $T_{cr}(\kappa)$ which resembles theoretical one found above within constant mobility scenario(see Fig.\ref{Fig5}c). In contrast to our model, both the thickness and the film conductivity(mobility) vary in the experiment. To account for both changes we use Eqs.(\ref{transcendental_equation}),(\ref{z(T)}) and the condition $z_{cr}(\kappa)=z(\xi)$ valid for arbitrary film thickness. Finally, we obtain threshold temperature for arbitrary sample size and disorder as it follows
\begin{equation}
T_{cr}= T_{F} \left[ \left(\frac{z_{cr}^{2}(\kappa)}{z_{m0}^{2}}\sigma(\kappa) \rho_{0}-1 \right )\frac{1}{\gamma} \right]^{1/2},
\label{T_cr}\\
\end{equation}
where $z_{m0}$,$\rho_{0}$ are constants deduced above for superconductor transition at $T_{cr}=0$. Note that the previous result for bulky sample specified by Eq.(\ref{T_c}) follows from Eq.(\ref{T_cr}) when $\kappa \rightarrow \infty$ and, then $z_{cr}(\kappa)=1$ and $\sigma \rho_{0}/z_{m0}^{2}=\mu/\mu_{\text{min}}$.

\begin{figure}[tbp]
\begin{center}\leavevmode
\includegraphics[width=1.0\linewidth]{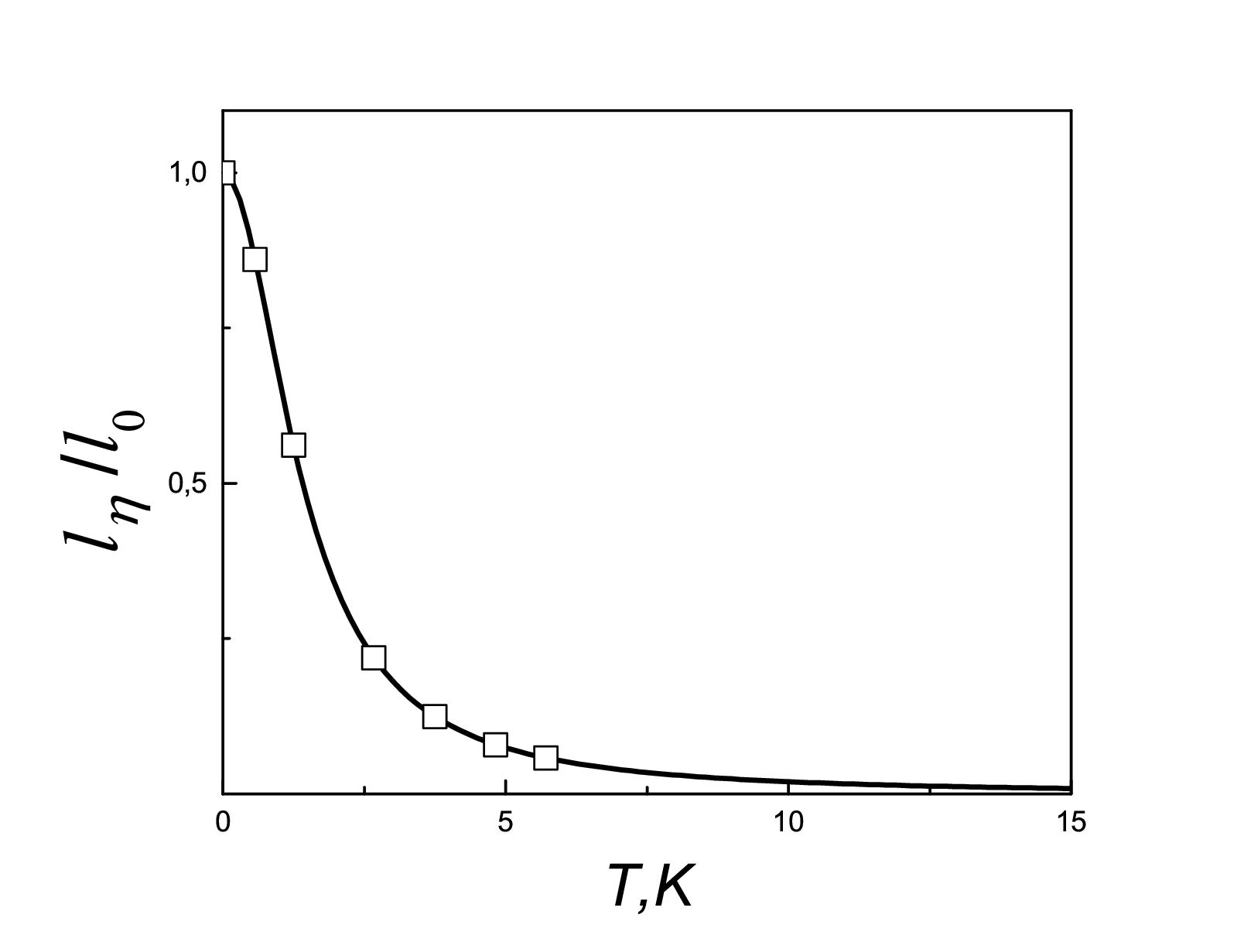} \caption[]{\label{Fig9} Dimensionless e-e relaxation time for amorphous bismuth vs
temperature ($l_{0}=18$A) extracted from experimental data\cite{Haviland89}. Square dots visualize the data in Fig.\ref{Fig8}
associated with superconductor transition at $T<T_{c}$.}
\end{center}
\end{figure}

With the help of Eq.(\ref{T_cr}) we are able to fit the experimental dependence $T_{cr}(\kappa)$ represented in Fig.\ref{Fig8b}a.
We use the dependence $\sigma(\kappa)$ in Fig.\ref{Fig8a}b and Eq.(\ref{T_cr}) keeping $\gamma/T_{F}^{2}$ as a fitting parameter.
Our best fit shown by the red curve Fig.\ref{Fig8b}a yields a parameter $\gamma/T_{F}^{2}=0.5 \text{K}^{-2}$. This finding help us to write down
the bismuth relaxation length as it follows
\begin{equation}
\l_{0}/l_{\eta}=1+0.5\cdot T^{2}
\label{visc_relax_time_real}\\
\end{equation}
valid for temperature range $T<T_{c}$. The dependence specified by Eq.(\ref{visc_relax_time_real}) is
extrapolated up to high temperatures $T\leq 14$K and, then plotted in Fig.\ref{Fig9}.

To illustrate the robustness of our approach, we compare the experimental dependence $R_{\square}(T_{cr})$ in Fig.\ref{Fig8a}a with that visualized by the red curve and followed from combined use of Eq.(\ref{T_cr}), dependence $\sigma(\kappa)$ and newly extracted parameter $\gamma$. The theory predictions are close to experimental data.

Let us now reproduce films resistance data set\cite{Haviland89} plotted in Fig.\ref{Fig8}b. Combining Eqs.(\ref{z(T)}),(\ref{SC_resistivity}),(\ref{visc_relax_time_real}) along with the relationships $\nu=2\kappa z(\xi)$,$z_{m}=\frac{z_{m0}}{\sigma(\kappa)\rho_{0}}$ we obtain the sheet resistance for certain layer thickness. In contrast to bottom curves related to thick films, the actual sample thicknesses for curves above the separatrix were not indicated in Ref.\cite{Haviland89}. To fix the situation, we compare the experimental value of the resistance $R_{\square}$ at $14$K for each film $d<d_{m}$ with that followed from our calculations and, then estimate the layer thickness roughly. The result of our efforts is shown in Fig.\ref{Fig8}b. As expected, the change from downturn $\frac{d\rho}{dT}>0$ to upturn $\frac{d\rho}{dT}<0$ behavior occurs when $\kappa=1$. Actually, the "insulating" behavior of the resistance for thin $\kappa<1$ samples follows from relationship $\frac{d\rho_{\eta}}{dT}\sim\frac{d\eta}{dT}\sim \frac{d\tau_{\eta}}{dT}<0$. The sets in both panels of Fig.\ref{Fig8} resemble one another excepting the shape and magnitude of the "insulating" curves above the separatrix. We attribute this discrepancy to possible hopping transport which would be relevant for Bi film of atomic length scale\cite{Komnik88}.

It is instructive to estimate the Bi-samples parameters. We remind you that there was no study of carrier density
in Ref.\cite{Haviland89}. In view of typical sample resistivity $\rho_{0}=410\mu\Omega \text{cm}$ close to that given by Ioffe-Regel
formalism\cite{Ioffe60} for dirty metal, we use this theory\cite{Gurvitch81} to calculate the carrier mean free path $l$. For a sample of size $d=7.2$ argued above to be a superconductor at $T_{cr}=0$ we obtain  $l=\frac{e^{2}Z^{2/3}}{3\hbar}\rho_{0}=9.9$A, where $Z=5$ is the number of valence electrons for bismuth. The present model gives the length of viscosity relaxation $l_{0}=\frac{5d_{m}^{2}}{4z_{m0}^{2}l}=18$A at $T=0$. While $l<l_{0}$, both lengths demonstrate the same order of magnitude\cite{Cheremisin21} and, hence support the applicability of the hydrodynamic approach in question.

\subsection{Phase diagram of superconductivity B(T) for massive specimen}

\label{Phase diagram B(T) of superconductivity for massive specimen}
The all previous discussion concerned the zero-current mode of the galvanic measurements. We demonstrated that growth of temperature
and(or) drop of the sample size destroys the superconductivity. In general, the enhancement of the applied currents and(or) magnetic fields is known\cite{Meissner33,Meissner33+,Silsbee17} to break down the superconductivity as well. Noticeably, the typical scale of critical magnetic fields ($\leq kG$) corresponds to classical limit $\Omega_{c}\tau\ll 1$ regarding the carrier transport. We now examine the influence of finite current and(or)
magnetic field on the superconductivity.

Our previous findings confirmed the crucial role of diamagnetic currents regarding the origin of superconductivity. Usually, the
diamagnetic currents are generally believed to be caused by cyclotron movement of electrons whose orbits are closed to (but contactless)
the inner sample walls. Noticeably, one must distinguish the diamagnetic contribution to susceptibility with paramagnetic one
caused by electron reflections from the sample walls, i.e so-called skipping trajectories.

\begin{figure}[tbp]
\begin{center}\leavevmode
\includegraphics[width=1.1\linewidth]{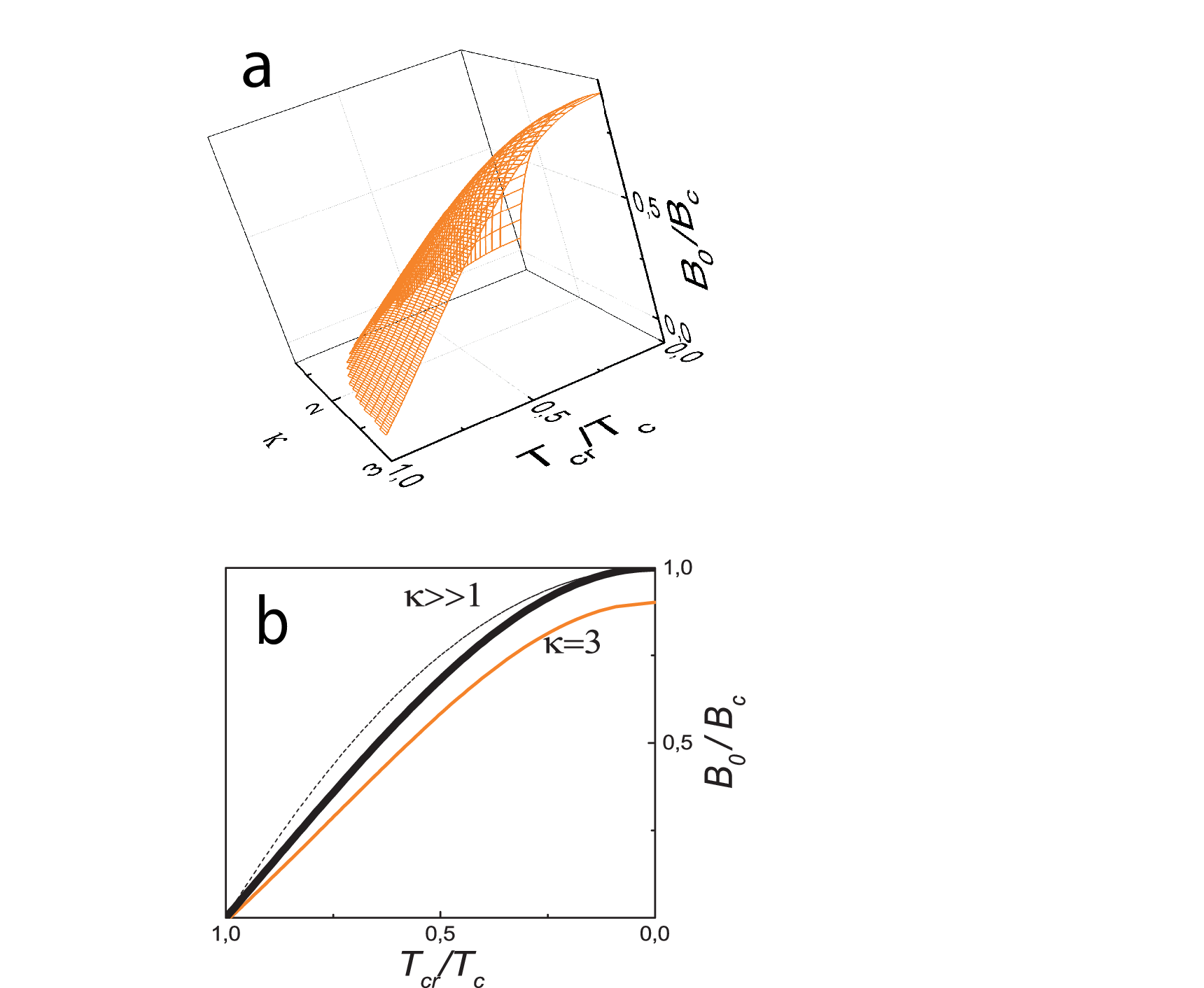} \caption[]{\label{Fig10} a) 3D plot of critical surface given by Eq.(\ref{Critical_condition_B+}) b) Phase diagram of superconductivity given by Eq.(\ref{Phase_diagram_theory}) for $z_{m}=0.8$ for massive($\kappa=\infty$) and finite size sample $\kappa=3$. The dashed curve depicts the empirical dependence specified by Eq.(\ref{Phase_diagram_empirical}).}
\end{center}
\end{figure}

Within classical range of the magnetic fields $\Omega_{c}\tau\ll 1$ one may imagine an electron moved along the cyclotron orbit during the mean
free time and, then experienced a collision with phonons and(or) impurities. Importantly, the curvature of cyclotron orbit trajectory counted for
subsequent collisions increases with magnetic field. For cyclotron trajectories of an electron in the vicinity of the top(bottom)
sample wall(see Fig.\ref{Fig1}) one must account for only the local magnetic field, namely $B_{0}$ in our notations. Evidence shows that the thickness
of the layer filled by diamagnetic current flux could be a linear function of transverse magnetic field, i.e. $\delta(1+bB_{0})$. Here, we write down a phenomenological constant $b$. Finally, we introduce a modified sample-size parameter
\begin{equation}
\kappa(B_{0}) = \frac{\kappa} {1+bB_{0}},
\label{kappa(B)}\\
\end{equation}
which depends on the transverse magnetic field $B_{0}$. The superconductivity threshold transition criteria given by Eq.(\ref{Critical_condition}) now yields
\begin{equation}
\kappa(B_{0}) \beta(\nu)=1.
\label{Critical_condition_B+}\\
\end{equation}
For the most important case of the massive specimen $\kappa \rightarrow \infty$, we find $\beta (2z\kappa)\simeq 1/z\kappa$.
Using Eq.(\ref{z(T)plus}) one obtains phase diagram of superconductivity
\begin{equation}
B_{0}=B_{c}\frac{\left(1+(z_{m}^{-2}-1)\Theta^{2}\right )^{-1/2}-z_{m}}{1-z_{m}}
\label{Phase_diagram_theory}\\
\end{equation}
where $B_{c}=(z^{-1}_{m}-1)/b$ is the critical magnetic field at zero temperature. As an example, the magnetic field driven
phase diagram specified by Eq.(\ref{Phase_diagram_theory}) for fixed $z_{m}=0.8$ is plotted in Fig.\ref{Fig10}b. For
$z_{m}\leq 1$ the phase diagram given by Eq.(\ref{Phase_diagram_theory}) can be fitted by quadratic dependence
\begin{equation}
B_{0}=B_{c}(1-\Theta^{2})
\label{Phase_diagram_empirical}\\
\end{equation}
usually reported to be a good approximation for the most of elementary superconductors. Note that in the vicinity of critical temperature
$T_{c}-T\ll T_{c}$ Eq.(\ref{Phase_diagram_theory}) gives a linear slope $\mid\frac{1}{B_{c}}\frac{dB_{0}}{d\Theta}\mid=(1+z_{m})z_{m} \leq 2$
close to that $\mid\frac{1}{B_{c}}\frac{dB_{0}}{d\Theta}\mid = 2$ provided by Eq.(\ref{Phase_diagram_empirical}).
This result is confirmed by numerous experiments.
Using Eq.(\ref{T_c}) one may find out the critical magnetic field at zero temperature
\begin{equation}
B_{c}=b^{-1}\left (\sqrt{1+\gamma(T_{c}/T_{F})^{2}}-1\right )
\label{Zero_Field}\\
\end{equation}
for massive samples. At low temperatures $\sqrt{\gamma}T_{c}/T_{F}\ll 1$ the critical temperature obeys quadratic dependence $B_{c}\sim \frac{\gamma}{2b}\left (\frac{T_{c}}{T_{F}}\right )^{2}$ while becomes linear in the opposite high temperature case, i.e. $B_{c}\sim \frac{\sqrt{\gamma}}{b}\frac{T_{c}}{T_{F}}$.

We emphasize that for arbitrary sample size Eq.(\ref{Critical_condition_B+}) defines the threshold surface in $T-d-B$ space.
The latter can be readily represented in terms of dimensionless sample size $\kappa$, then the critical temperature $\Theta_{cr}$ and magnetic field $B_{0}/B_{c}$ related to massive specimen values. The 3D plot is shown in Fig.(\ref{Fig10})a for fixed $z_{m}=0.8$ and finite sample size $\kappa<3$. As expected, for massive sample, i.e. when $\kappa \rightarrow \infty $ the critical diagram $B_{0}(T)$ approaches that depicted in Fig.(\ref{Fig10})a.

\subsection{Magnetic field turned the normal phase to superconductivity transition}
\label{Magnetic field turned the normal phase to superconductivity transition}
\begin{figure}[tbp]
\begin{center}\leavevmode
\includegraphics[width=1.0\linewidth]{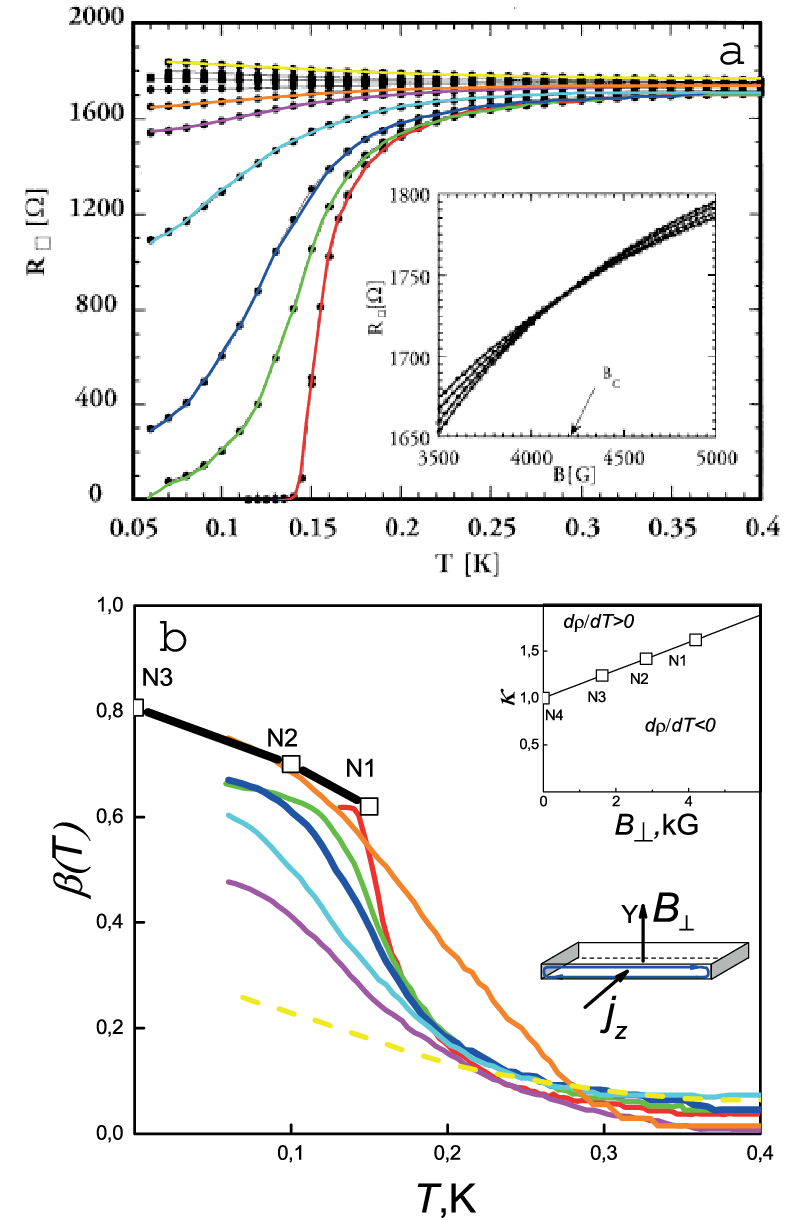} \caption[]{\label{Fig11} a) Sheet resistance for Mo$_{21}$Ge sample N1( $d=80$A )
under Ref.\cite{Kapitulnik95} for $B_{\perp}=0,0.5;1;2;3;4;4.4;4.5;5.5;6$ kG.(from bottom to top) b) Universal function $\beta$ vs
temperature deduced from each of the colored curves in panel a. Inset: (top) dependence specified by Eq.(\ref{R(T)_behavior_criterion}); (bottom) Experimental setup\cite{Kapitulnik95}. Black squares represent the data for samples denoted in Table\ref{tabl1}.}
\end{center}
\end{figure}

Recall that the result of Sec.\ref{Sample-size driven normal state to superconductivity transition} concerned the change from upturn $\frac{d\rho}{dT}> 0$ to downturn $\frac{d\rho}{dT}< 0$ behavior of the sheet resistance as the sample thickness decreased. At low currents this happens for certain sample size when $\kappa=1$. Obviously, the current enhancement leads to subsequent growth of the transverse magnetic field( see Fig.\ref{Fig6}a). Remind that the strength of diamagnetic currents at the upper(down) sample walls is linear to local transverse field $B_{0}$. Hence, the derivative $\frac{d\rho}{dT}$ would change the sign when $\kappa=1+bB_{0}$. We conclude that for fixed slab thickness the enhanced current could result in the resistance behavior similar to that seen in Fig.\ref{Fig8}a. By now, we unaware of any experiment dealing with the proposed effect. In contrast, a film placed into perpendicular magnetic field $B_{\perp}$
demonstrates the magnetic field driven change of sheet resistance shape\cite{Hebard90,Kapitulnik95,Markovic98}. The experimental setup is shown in Fig.\ref{Fig11}b,inset. The typical resistance set measured for different fields is reproduced in Fig.\ref{Fig11}a under Ref.\cite{Kapitulnik95}. We now intend to examine this effect.

We argue that the applied perpendicular field $B_{\perp}$ produces the tangential component of the magnetic field $B_{y}\parallel B_{\perp}$ at the
sample side faces shown by grey color in Fig.\ref{Fig11}b, inset. Indeed, let us assume the sample as a thin disk which,
in turn, can be viewed as limiting case of the rotation ellipsoid. In this case the search of the induced magnetic field yields
the textbook\cite{Rose-Innes78} result $B_{y}=\frac{B_{\perp}}{1-N}$, where $0<N<1$ is so-called demagnetization factor.
For thin disk $N\simeq 1$. Recall that whenever the tangential component of the magnetic field is present at arbitrary sample surface
the superconductivity becomes suppressed. We speculate that the component $B_{y}$ present at the grey face of the film side plays the role of magnetic field $B_{0}$ embedded into Eq.(\ref{kappa(B)}). As a result, we obtain the modified criteria
\begin{equation}
\kappa= 1+b^{*}B_{\perp}.
\label{R(T)_behavior_criterion}\\
\end{equation}
for derivative $\frac{d\rho}{dT}$ change instead of previous one $\kappa=1$ valid in absence of the perpendicular magnetic field.
Here $b^{*}=b/(1-N)$ is the sample dependent coefficient. The linear dependence specified by Eq.(\ref{R(T)_behavior_criterion})
is reproduced in Fig.\ref{Fig11}b,inset. Remarkably, a similar dependence was reported in experiment\cite{Markovic98}.

We argue that separatrix of the sheet resistance curves in Fig.\ref{Fig11}a occurs for sample N1 of thickness $d=80$A,
the sheet resistivity $R_{cr}=1750\Omega$ in presence of magnetic field $B_{\perp}=4.19$kG. Then, at zero field the resistance data
demonstrate the superconductor transition at $T_{cr}=0.15$K. Similar studies\cite{Kapitulnik95} reported for thinner($d=70$A) sample N2 give the respective
values collected in Table \ref{tabl1}. For both samples the normal resistivity is the same, namely $\rho=1400\mu\Omega \text{cm}$ at $T=0.14$K,
indicating a similar disorder in both cases. Therefore, we use the basic results of our model implying a constant disorder strength.
Using the parameters of samples N1,2 and Eq.(\ref{R(T)_behavior_criterion}) we define the minimal size value $d_{m}=50$A and coefficient $b^{*}=1.48\cdot 10^{-4}{\text G}^{-1}$. We calculate the reduced size $\kappa=d/d_{m}$ for each sample and, then use the critical diagram in Fig.\ref{Fig5}a
to find out parameter $z$ and value of the function $\beta(2\kappa z)$ respectively. The results are archived in Table \ref{tabl1}.

\begin{table}
\caption{\label{tabl1}Parameters of the Mo$_{21}$Ge samples N1,2 studied in Ref.\cite{Kapitulnik95} and predicted(N3,4) by present theory.}
\begin{tabular}{@{}*{8}{l}}
\hline\hline
       Samp.&$d[A]$&$\kappa$&$B_{\perp}$[kG]&$R_{cr}$[$\Omega$]&$T_{cr}$[K]&$z(T_{cr})$&$\beta(T_{cr})$\\
\hline\hline
       N1&80&1.62&4.19&1750&0.15&0.89&0.62 \\
       N2&70&1.42&2.82&2026&0.1&0.83&0.70 \\
\hline
       N3&61&1.24&1.62&-&0&0.70&0.81 \\
       N4&50&1&0&-&-&0&- \\
\hline\hline
\end{tabular}
\end{table}

Further analysis of the resistance data for sample N1 in Fig.\ref{Fig11}a gives evidence of superconductor transition at $T_{cr}=0$ which would occur at a certain magnetic field $B_{\perp}\simeq 1.1$kG. This case of the special interest providing disclosure of key parameter $z_{m}$ of our model. Indeed, for reduced film size $\kappa=1.62$ of sample N1 the solution of Eqs.(\ref{Critical_condition_B+}),(\ref{kappa(B)}) under the substitution $bB_{0}\rightarrow b^{*}B_{\perp}$ yields the parameter $z=z_{m}=0.7$. Then, we find $\kappa_{m}=1.24$ with the help of critical diagram shown in Fig.(\ref{Fig5})a. We conclude that superconductivity transition at $T_{cr}=0$ and zero magnetic field would be feasible for hypothetical sample of thickness $\kappa_{m}d_{m}=61$A labeled as N3 in Table\ref{tabl1}. According to Eq.(\ref{kappa(B)}) the sample N3 would exhibit a change in metallic vs "isolating" behavior of resistance
at $B_{\perp}=1.62$kG. Also, we collect in Table\ref{tabl1} the parameters for even thinner hypothetic sample N4 of thickness $d=d_{m}=50$A which
would exhibit a constant resistance $\rho=\rho_{D}$ at zero magnetic field( see Fig.\ref{Fig9}b, inset).

Recall that resistivity specified by Eq.(\ref{SC_resistivity}) depends on the universal function $\beta(T)$. For each curve
in Fig.\ref{Fig11}a the function $\beta(T)$ can be extracted. One take an interest whether the all curves can be scaled by the same dependence $\beta(T)$.
Taking into account the modified parameter specified by Eq.(\ref{kappa(B)}) and Eq.(\ref{SC_resistivity}) we calculate and, then plot in Fig.\ref{Fig11}b
the result. In contrast to rough resistance data ranges from zero to values of the order of $\sim k\Omega$, the range of function $\beta(T)$ magnitude falls
into much narrow range $0.3<\beta(T)<0.8$ seen in Fig.\ref{Fig11}b. For completeness, we add in the same plot the dependence $\beta(T_{cr})$
followed from Table\ref{tabl1} for samples N1-3 when the magnetic field is zero. In conclusion, the present analysis of the experimental data\cite{Kapitulnik95} provides strong support in favor of our model.

\section{Conclusions}
In conclusion, we discover the self-consistent Hall Effect in a bar conductor taking into account both the diamagnetism and finite viscosity
of 3D electron liquid. We demonstrate that under certain condition the resistivity of the sample vanishes exhibiting the transition
to superconductivity. The current is pinched nearby the inner sample walls while the magnetic is pushed out from the sample bulk.
Within low current limit the threshold temperature of superconductivity is calculated for arbitrary carrier dissipation and the sample size.
For sample size and(or) carrier mobility being lower than a certain minimum values the superconductivity state cannot be realized.
Sample-size and magnetic field driven transition from zero-resistance state to normal state is compared with experimental data.
Phase diagram in terms of threshold temperature vs applied current and(or) magnetic field is calculated. The temperature dependence of
the e-e scattering time for amorphous bismuth is extracted from the experimental data for the first time.
\bibliography{SC_2D}

\end{document}